\documentclass[bibyear]{aa}
\makeatletter
\renewcommand*\aa@pageof{, page \thepage{} of \pageref*{LastPage}}
\makeatother
\usepackage{url}
\usepackage{natbib}
\usepackage{multicol}
\usepackage{caption}
\usepackage{subcaption}
\usepackage{animate}
\bibpunct{(}{)}{;}{a}{}{,}
\usepackage{graphicx}
\usepackage{txfonts}
\usepackage[dvipsnames]{xcolor}
\usepackage{amsmath}
\usepackage{hyperref}
\hypersetup{
    colorlinks=true,
    citecolor=blue,
    linkcolor=blue,
    filecolor=magenta,      
    urlcolor=cyan}
\usepackage{caption}
\usepackage{lipsum}
\usepackage{media9}
\usepackage[normalem]{ulem}

\def\nhism{{$N_{\rm H,ISM}$}}
\def\cm{cm$^{-2}$}

\begin{document}

\title{X-ray properties and obscured fraction of AGN in the J1030 \textit{Chandra} field}
   %\subtitle{I.}
\titlerunning{X-ray properties and obscured fraction of AGN in the J1030 \textit{Chandra} field}

\author{Matilde Signorini\inst{1,2,3}\thanks{\email{matilde.signorini@unifi.it}} \and Stefano Marchesi\inst{4,5,6} \and Roberto Gilli\inst{4} \and  Marcella Brusa\inst{4,6} \and Andrea Comastri\inst{4} \and Quirino D'Amato\inst{11,12} \and Kazushi Iwasawa\inst{8,9} \and Giorgio Lanzuisi\inst{4} \and Giovanni Mazzolari\inst{4} \and Marco Mignoli\inst{4}  \and Alessandro Peca\inst{7}    \and Isabella Prandoni\inst{12} \and Paolo Tozzi\inst{2} \and Cristian Vignali\inst{6,4} \and Fabio Vito\inst{4} \and Colin Norman\inst{10} } 

\institute{Dipartimento di Fisica e Astronomia, Universit\`a di Firenze, via G. Sansone 1, 50019 Sesto Fiorentino, Firenze, Italy
         \and
             INAF - Osservatorio Astrofisico di Arcetri, Largo Enrico Fermi 5, I-50125 Firenze, Italy
        \and 
            University of California-Los Angeles, Department of Physics and Astronomy, 430 Portola Plaza, Los Angeles, CA 90095-1547, USA
        \and
            INAF - Osservatorio di Astrofisica e Scienza dello Spazio di Bologna, Via Piero Gobetti, 93/3, 40129, Bologna, Italy
        \and 
            Department of Physics and Astronomy, Clemson University, Kinard Lab of Physics, Clemson, SC 29634, USA
        \and
            Dipartimento di Fisica e Astronomia, Università degli Studi di Bologna, via Gobetti 93/2, 40129 Bologna, Italy
        \and 
            Department of Physics, University of Miami, Coral Gables, FL 33124, USA
        \and
            Institut de Ciències del Cosmos (ICCUB), Universitat de Barcelona (IEEC-UB), Martí i Franquès, 1, 08028, Barcelona, Spain
        \and 
            ICREA, Pg. Luís Companys 23, 08010 Barcelona, Spain
        \and
            Space Telescope Science Institute, 3700 San Martin Drive, Baltimore, MD 21218, USA
        \and 
             Scuola Internazionale Superiore di Studi Avanzati (SISSA), Via Bonomea 265, 34136 Trieste, Italy
        \and 
            Istituto di Radioastronomia (IRA), Via Piero Gobetti 101, 40129 Bologna, Italy
}

\abstract

\abstract{The 500ks \textit{Chandra} ACIS-I observation of the field around the $z=6.31$ quasar SDSS J1030+0524 is currently the 5th deepest extragalactic X-ray survey. The rich multi-band coverage of the field allowed for an effective identification and redshift determination of the X-ray source counterparts: to date a catalog of 243 extragalactic X-ray sources with either a spectroscopic or photometric redshift estimate in the range $z\approx0-6$ is available over a 355 arcmin$^2$ area. Given its depth and the multi-band information, this catalog is an excellent resource to investigate X-ray spectral properties of distant Active Galactic Nuclei (AGN) and derive the redshift evolution of their obscuration. We performed a thorough X-ray spectral analysis for each object in the sample, measuring its nuclear column density $N_{\rm H}$ and intrinsic (de-absorbed) 2-10 keV rest-frame luminosity, $L_{2-10}$. Whenever possible, we also used the presence of the Fe K$_\alpha$ emission line to improve the photometric redshift estimates. 
\\We measured the fractions of AGN hidden by column densities in excess of 
$10^{22}$ and $10^{23}$cm$^{-2}$ ($f_{22}$ and $f_{23}$, respectively) as a function of $L_{2-10}$ and redshift, and corrected for selection effects to recover the intrinsic obscured fractions. At $z\sim 1.2$, we found $f_{22}\sim0.7-0.8$ and $f_{23}\sim0.5-0.6$, respectively, in broad agreement with the results from other X-ray surveys. No significant variations with X-ray luminosity were found within the limited luminosity range probed by our sample (log$L_{2-10}\sim 42.8-44.3$). \\When focusing on luminous AGN with log$L_{2-10}\sim44$ to maximize the sample completeness up to large cosmological distances, we did not observe any significant change in $f_{22}$ or $f_{23}$ over the redshift range $z\sim0.8-3$. Nonetheless, the obscured fractions we measure are significantly higher than what is seen in the local Universe for objects of comparable intrinsic luminosity, pointing towards an increase of the average AGN obscuration towards early cosmic epochs, as also observed in other X-ray surveys. \\  
We finally compared our results with recent analytic models that ascribe the larger obscuration observed in AGN at high redshifts to the dense interstellar medium (ISM) of their hosts. When combined with literature measurements, our results favor a scenario in which both the total column density of the ISM and the characteristic surface density of its individual clouds increase towards early cosmic epochs as \nhism$\propto(1+z)^{\delta}$, with $\delta\sim 3.3-4$, and $\Sigma_{c,*}\propto(1+z)^2$, respectively.
}

   \date{Received XX; accepted XX}

    
\maketitle

\section{Introduction}\label{sec:intro}
The characterization of Active Galactic Nuclei (AGN) demographics and their evolution is crucial to understand the history of the accretion onto supermassive black holes (SMBHs) and their relation with their host galaxies. \\
We know that the masses of SMBHs residing in the centers of most galaxies correlate with the host properties, such as the stellar luminosity, the stellar mass, and the bulge velocity dispersion \citep[e.g.,][]{Marconi03, Ferrarese06, Kormendy13, DeNicola19}. These correlations indicate a co-evolution scenario of SMBHs and galaxies across cosmic epochs that has been observationally investigated and theoretically modeled \citep[e.g.,][]{Croton06, Somerville08, Fabian12, Habouzit19, Ricarte19}, but is still far from being understood in its entirety. For example, in the early Universe, the very formation and accretion processes leading to the first SMBHs are still debated. A simple accretion history on stellar mass black holes formed by the first stars is challenged by the discoveries of SMBHs of 1-10 billion of solar masses at redshifts higher than 6 (\citealp{Mortlock11, Wu15, Banados16, Farina22}; \citealp[see also][for a recent review ]{Fan22}). To match these masses, the accretion process needs to be Eddington-limited or even super-Eddington for long times, or we need to have very massive black holes "seeds" to start with. \\Although both the accretion process and the masses of the "seeds" are still debated, the majority of galaxies are thought to have undergone an active nuclear phase, in which they can be detected as AGN \citep{Kormendy13}. This makes investigations of AGN at different cosmic epochs crucial to understand the growth and evolution of both SMBH and galaxies. 

However, the presence of gas and dust, both in the innermost nuclear regions and across the whole host galaxy, poses a significant challenge to AGN detection and characterization, given the damping of the %absorption/extinction of
emission in the optical/UV band, where AGN intrinsic power peaks. AGN population synthesis models agree that most SMBH growth is hidden to our view by large gas column densities \citep[see, e.g.,][]{Gilli07, Ueda14, Ananna19}. Such a scenario has been confirmed by several observational works \citep[e.g.,][]{Lanzuisi18, Vito18}, that further show that, at high redshifts ($z>$3--4), the fraction of luminous AGN obscured by column densities N$_{\rm H}>$10$^{23}$\,cm$^{-2}$ is particularly high, $\sim$80\,\%, as opposed to 20--30\,\% measured in the local Universe \citep[see, e.g.,][]{Torres21}.

X-ray surveys provide one of the most effective ways to detect obscured AGN over a wide range of redshifts and luminosities \citep[see, e.g.,][for an extensive review]{Brandt15, Xue17, Hickox18}, and are therefore key to finding and characterizing accreting supermassive black holes in the early Universe. Indeed, while the AGN emission in the X-rays is $<$10\,\% of the overall AGN luminosity \citep[e.g.,][]{Lusso_12, Duras20}, it suffers very little contamination from to non-AGN processes (e.g., X-ray binaries, diffuse gas emission), and is significantly less biased against obscuration than optical emission. These reasons make X-ray surveys a great and efficient way to detect AGN and to characterize them and their obscuration. 
Several works have used X-ray data to investigate the evolution of AGN obscuration with cosmic times \citep{Lafranca05, Tozzi06, Treister06, Ueda14, Buchner15, Aird15, Liu_17, Vito18, Lanzuisi18, Iwasawa_20, Peca23}. Such works find that the fraction of obscured objects increases with redshift, but the physical origin of this trend is not completely understood \citep[see, e.g.][]{Iwasawa12}. 
There are, indeed, arguments to suggest that the properties of the obscuring torus do not evolve significantly: for example, the Spectral Energy Distributions (SED) of AGNs are the same at very different redshifts \citep{Richards06, Bianchini19}. This means that the covering factor of the dusty torus is unlikely to change with time. The properties of the interstellar medium (ISM) of the host galaxies, instead, vary significantly with time. The content of gas is higher at early cosmic times \citep[see, e.g.,][]{Scoville17,Tacconi18,Aravena20}, and galaxies are also smaller in size \citep{Allen17, Fujimoto17}. This means that, as the ISM density increases at high redshift, its column density can reach very high values and be the principal contribution to the obscuration of AGN, as shown in several recent works \citep[e.g.,][]{Circosta19,Damato20,Gilli22}. This has been shown also by hydrodynamic \citep{Trebitsch19} and cosmological \citep{Ni20} simulations.

In this work, we investigate the X-ray properties and derive the obscured fraction of the AGN sample in the J1030 \textit{Chandra} deep survey. In 2017, \textit{Chandra} observed a 355 arcmin$^2$ region around the z = 6.31 quasar SDSS J1030+0525 for $\sim$500 ks. The field around it has dense multi-wavelengths coverage, being observed with MUSYC-DEEP, HST/WFC3, HST/ACS, VLT/MUSE, WIRCam, IRAC \citep[see, e.g.][]{Peca_21}. 
The \textit{Chandra} survey has a 0.5-2 keV flux limit f$_{0.5-2 keV} = 6\times10^{-17}$erg s$^{-1}$cm$^{-2}$ in the central square arcmin and it is, to date, the fifth deepest extragalactic X-ray field \citep{Nanni_20}. The survey resulted in the detection of 256 sources, of which 7 are identified as stars, based on their spectra (three of them), or on their brightness in the K band and low X-ray-to-optical rate \citep{Nanni_20, Marchesi_21}. Among the remaining 249 extragalactic sources, \cite{Marchesi_21} were able to compute a photo-z for 243 of them, which make the sample considered in this work.\\
Multiple spectroscopic campaigns allowed the determination of the spectroscopic redshifts for 135 objects out of these 243 \citep[i.e., 56\,\% of the extragalactic sample,][]{Marchesi_21, Marchesi23}. Here, we present the complete spectral analysis of the X-ray spectra of the 243 \textit{Chandra} J1030 extragalactic objects. Our goal is to determine the physical properties of these sources and study the evolution of the obscured AGN fraction with luminosity and redshift. The paper is structured as follows: in Section~\ref{sec:reduction}, we describe the X-ray sample and the reduction of the \textit{Chandra} data; in Section~\ref{sec:analysis}, we describe the X-ray spectral analysis and its overall results for the sample; in Section~\ref{sec:obscured_fraction} we derive the obscured AGN fraction in the J1030 \textit{Chandra} field as a function of luminosity and redshift and for different absorption threshold; in Section~\ref{sec:discussion} we discuss the results, the physical interpretations and the limits of our work, and in Section~\ref{sec:conclusions} draw our conclusions. Through the rest of the work, we assume a flat $\Lambda$CDM cosmology with $H_0=69.6 $ km s$^{-1} $Mpc$^{-1}$, $\Omega_m=0.29$ and $\Omega_{\Lambda}=0.71$ \citep{Bennett14}.

\section{Sample description and X-ray spectral extraction}\label{sec:reduction}
\begin{figure}
\centering
\includegraphics[width=0.99\linewidth]{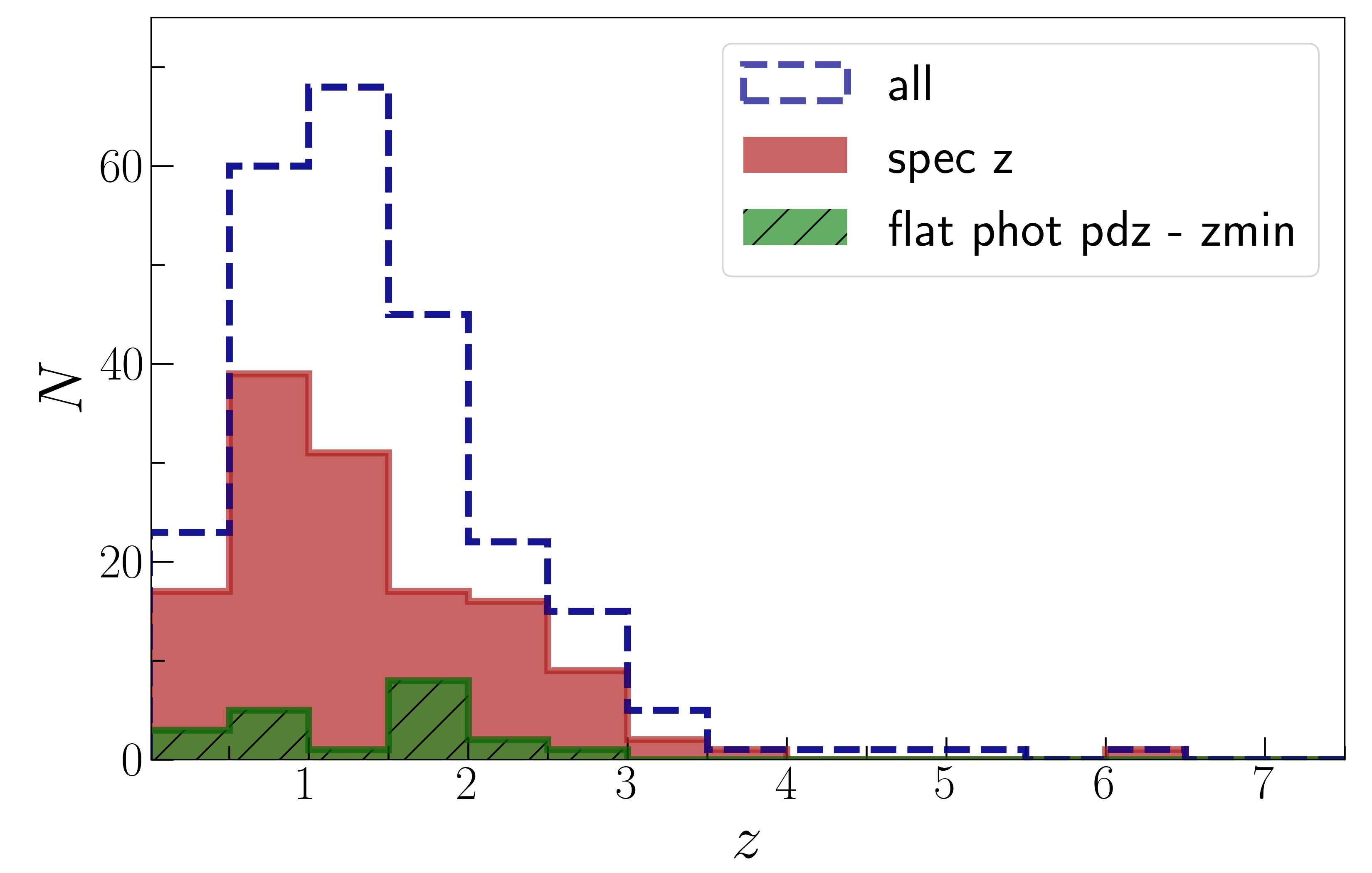}
\caption{Redshift distribution of the J1030 \textit{Chandra} catalog. In blue dashed, the distribution for the whole sample. In red filled, the distribution for the objects with spectroscopic redshift. A spectroscopic redshift is available for 135 out of 243 objects. In green striped, we show the redshift lower limit for the 20 objects for which the photometric redshift probability distribution is flat.}
\label{fig:z_hist}
\end{figure}

The \textit{Chandra} J1030 extragalactic sample is made of 243 objects, for which we have a redshift estimate that can be either photometric or spectroscopic. In Figure \ref{fig:z_hist} we show the redshift distribution of the objects in the catalog; a spectroscopic redshift estimate is available for 135 of them. For 20 out of the 108 photometric ones, the redshift probability distribution is flat \citep[see ][]{Marchesi_21}\footnote{The photometric SED and redshift probability distributions can be found on the website: \url{http://j1030-field.oas.inaf.it/xray_redshift_J1030.html}}. In Figure \ref{fig:z_hist} we show their minimum redshift estimate. The 135 objects with spectroscopic redshift also have spectral classification and are divided into four categories \citep[see][]{Marchesi_21}: 20 Narrow-Line AGNs (NL-AGN), 43 Broad-Line AGNs (BL-AGN), 32 Early Type Galaxies (ETG), and 40 Emission Line Galaxies (ELG). These numbers above are updated with respect to the one reported in \citet{Marchesi_21}, following new spectroscopic observations \citep{Marchesi23}.\\
Regarding the objects with only photometric redshift, recent works have proposed a way to derive an additional redshift estimate from the X-ray spectra \citep[e.g. ][]{Simmonds18, Sicilian22}. The method was tested for a subsample of the catalog, in \cite{Peca_21}. This method, however, requires highly obscured objects with a high number of counts ($N>150$, in \cite{Sicilian22}) to give redshift estimates that are more accurate and reliable than the photometric ones. Given the average properties of the sources in our sample, the X-ray spectrum is likely to provide a more precise redshift estimate only when the Fe K$\alpha$ line is detected. This will be discussed in  Section \ref{sec:analysis}.

\begin{figure}
\centering
\includegraphics[width=0.99\linewidth]{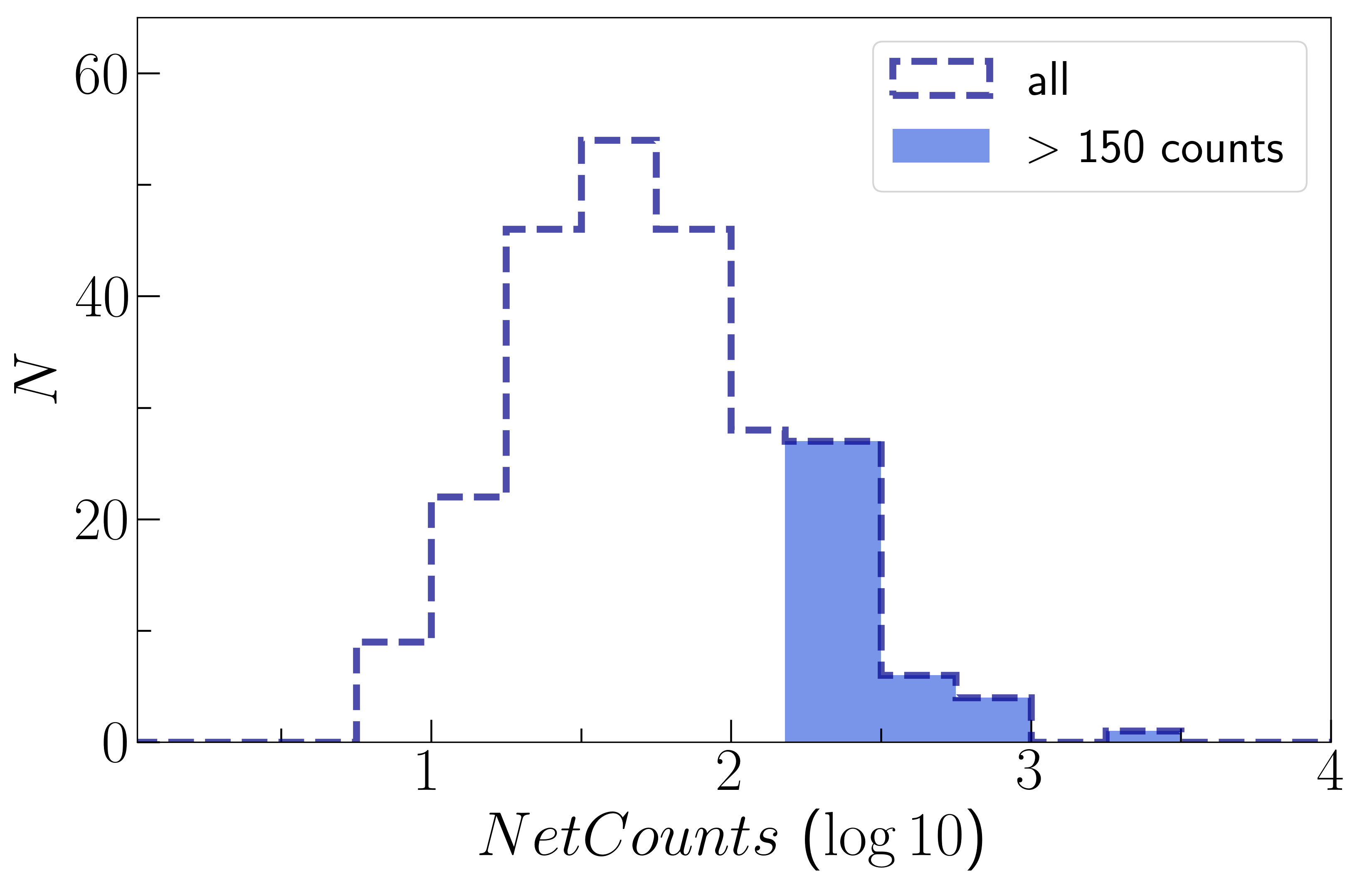}
\caption{Counts histogram for the J1030 field \textit{Chandra} catalog. There are  39 out of 243 objects which have more than 150 net counts in the 0.5-7\,keV range, shown as the light blue filled part of the histogram.}
\label{fig:cnts_hist}
\end{figure}

The spectra are extracted using the software \textit{Chandra} Interactive Analysis of Observations (CIAO) v.4.13. For the choice of the extraction radius, we performed a preliminary ad-hoc analysis. We argue that, as the PSF broadens with the increasing of the off-axis angle of the object, the best choice for the extraction radius might be different for objects at different off-axis angles. Furthermore, we expect to include more signal in a larger radius when the Signal to Noise Ratio (S/N) is higher; given this, we might need different radii between low and high-count objects.

To investigate this issue, we performed an analysis on a randomly chosen subsample of 35 objects which spans the off-axis - counts plane in the same way as the whole sample. We extracted and fitted the spectra obtained with different extraction radii, corresponding to the 75\% 80\%, 85\%, 90\%, and 95\% of the Encircled Energy. We then compared the S/N obtained with each Encircled Energy choice. The S/N does vary significantly between the different choices and, more importantly, there is no clear trend of the maximum of the S/N with the off-axis angle and/or the object counts. Therefore, we deemed an extraction radius R that corresponds to the 90\% of the Encircled Energy as a good choice for all the objects in our sample, consistent with what is already present in the literature \citep[e.g.,][]{Marchesi16}.

For each object, we also extracted a background spectrum in an annulus of radii R+2.5$^{\prime \prime}$ and R+20$^{\prime \prime}$, manually removing from the annulus any possible contaminating source. The selected background regions provided a sufficient sample for background estimation, allowing for the spectral fitting analysis to proceed. For each object, we used the CIAO command \textit{specextract} to extract the source and the background spectrum and to build the response matrix (RMF) and the ancillary response files (ARF). This was done for each of the 10 observations and the results were then combined with the CIAO tool $combine\_spectra$. To avoid empty channels, the resulting spectra were binned to a minimum of one count per bin. In the end, for each object, we produced the combined source spectrum, the combined background spectrum, and the combined RMF and ARF files. 

\section{Spectral analysis}\label{sec:analysis}
Once the spectra and the ancillary files were derived, we fitted them using  \textit{sherpa} \citep{sherpa}, fitting the background together with the source. We further discuss the background fitting procedure in Appendix~\ref{sec:app_bkg_fit}. 

The source spectral shape is modeled with an absorbed power-law. The Galactic absorption ($N_{\rm H,Gal}$ = 2.5$\times$10$^{20}$\,cm$^{-2}$) and the redshift are fixed parameters, while the column density at the source redshift, $N_{\rm H}$, is always assumed to be a free parameter. In principle, the power law photon index, $\Gamma$, should also be left free to vary. However, given the well-known degeneracy between $\Gamma$ and N$_{\rm H}$, in low-statistic spectra, a fit with both parameters free to vary can lead to unreliable results. For this reason, we decided to fix the photon index $\Gamma$ and leave the column density $N_{\rm H}$ as the only free parameter in sources with 0.5--7\,keV net counts below a given threshold. After some tests, we decided to put the threshold at 150 net counts in the 0.5--7\,keV band. In Figure~\ref{fig:cnts_hist} we show the net counts distribution (in the 0.5-7 keV band) for the objects in the catalog. Given the Poissonian nature of the data, we used the C statistic to perform the fit \citep{Cash79}. 

\begin{figure}
\centering
\includegraphics[width=0.99\linewidth]{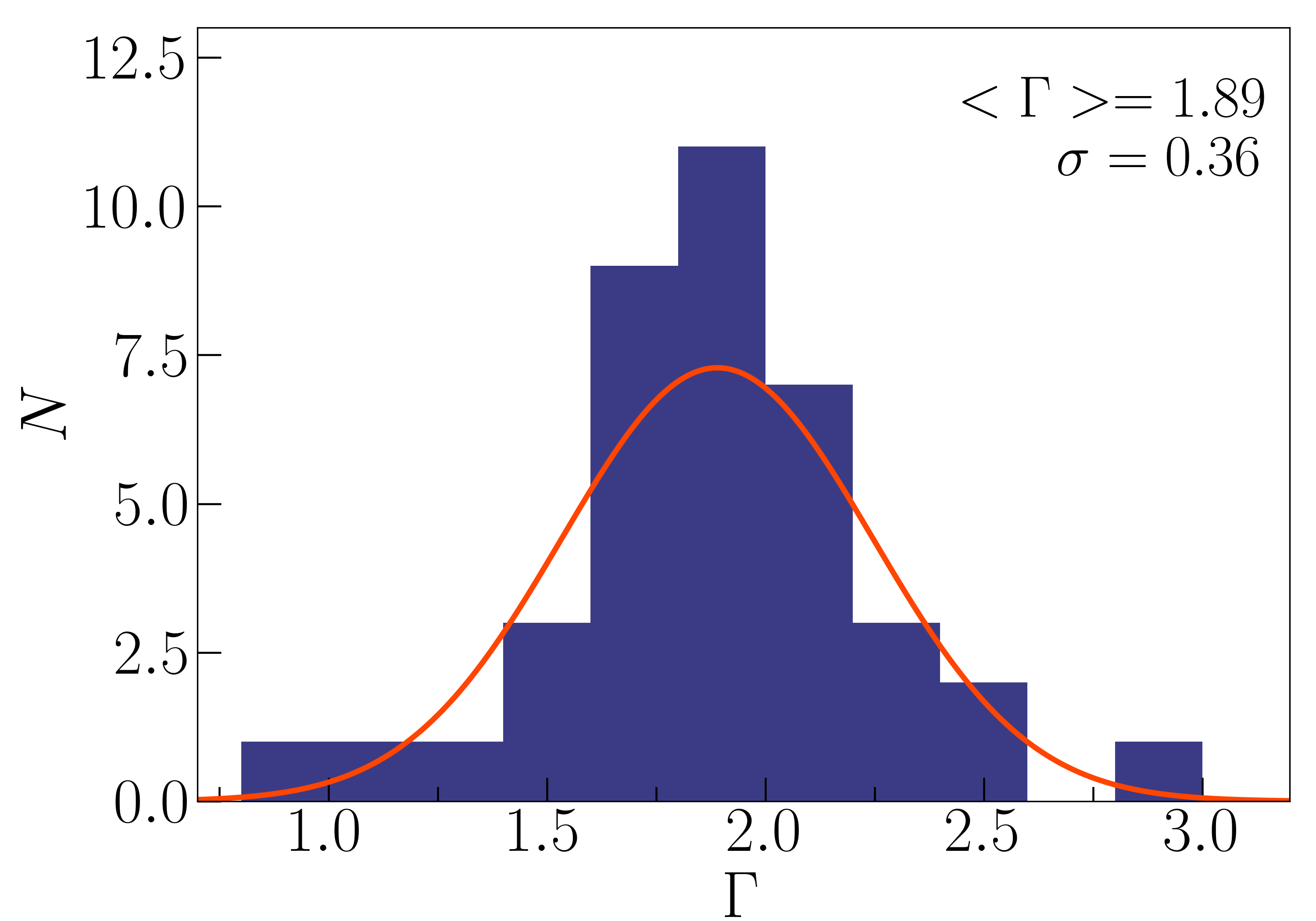}
\caption{Photon index distribution for the 39 objects with more than 150 net counts. A Gaussian fit gives $\langle\Gamma\rangle = 1.89, \sigma=0.36$.}
\label{fig:gamma_hist}
\end{figure}

\begin{figure*}
\centering
\begin{subfigure}{.49\textwidth}
  \centering
  \includegraphics[width=.99\linewidth]{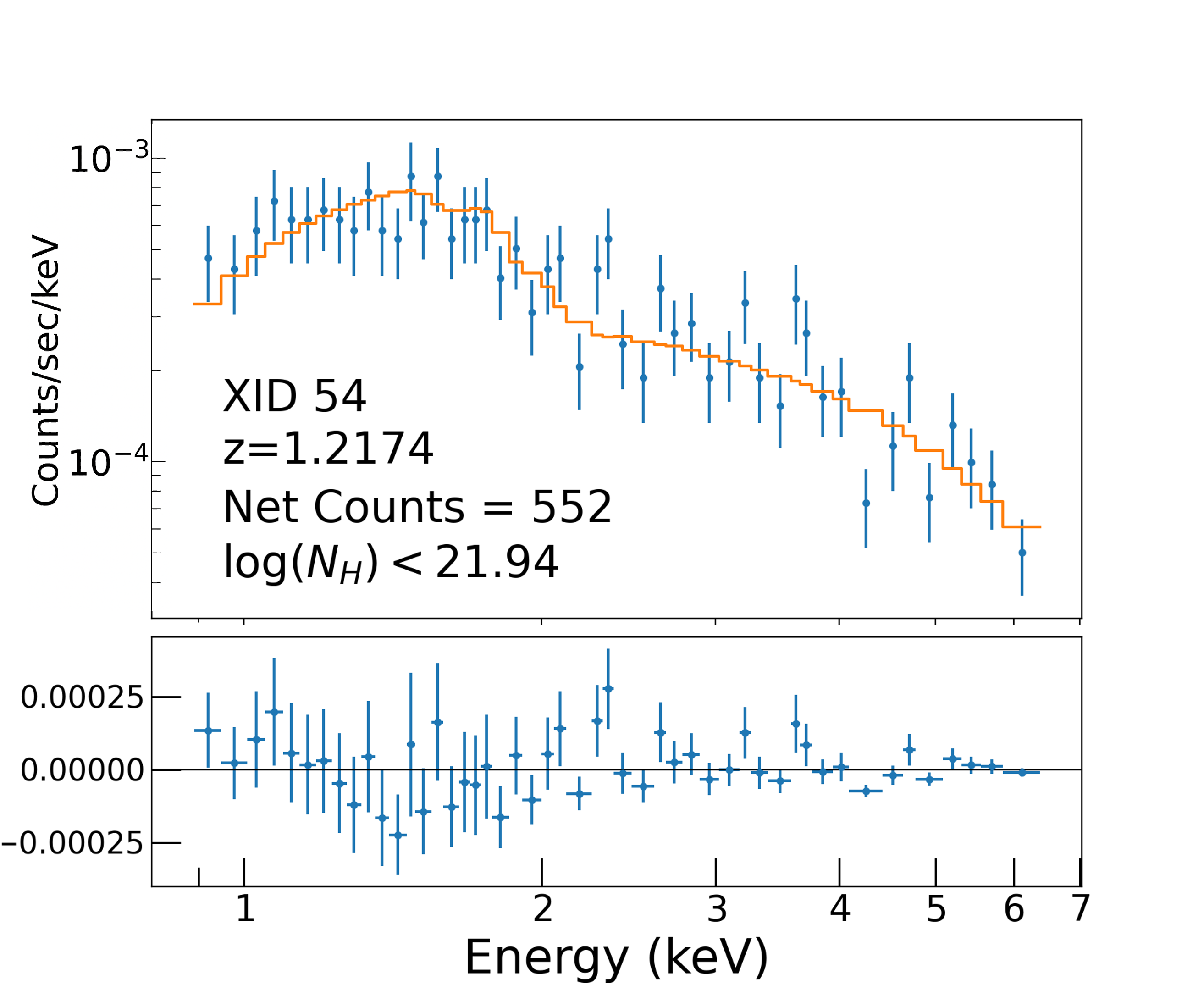}
\end{subfigure}%
\begin{subfigure}{.49\textwidth}
  \centering
  \includegraphics[width=.99\linewidth]{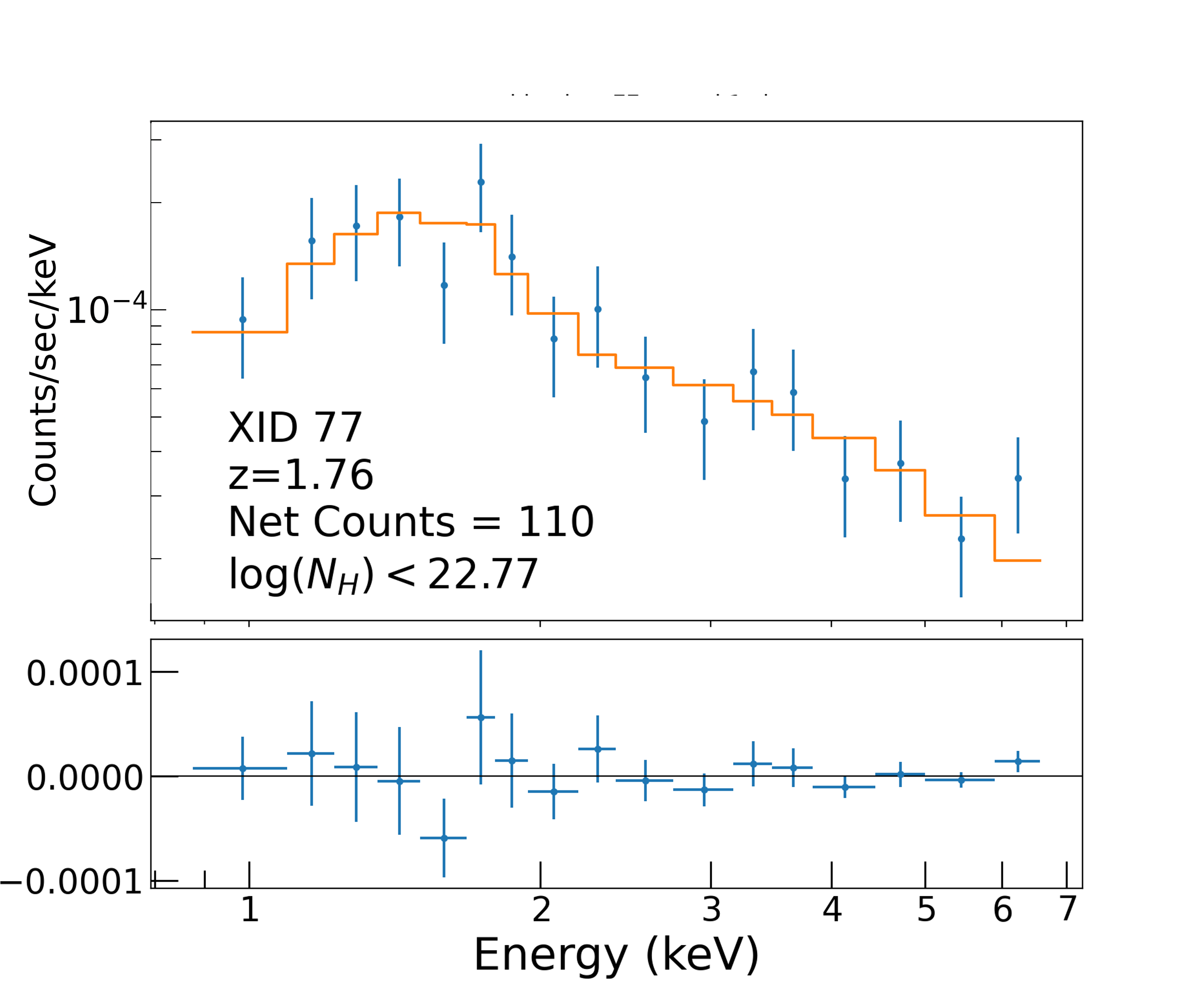}
\end{subfigure}
\begin{subfigure}{.49\textwidth}
  \centering
  \includegraphics[width=.99\linewidth]{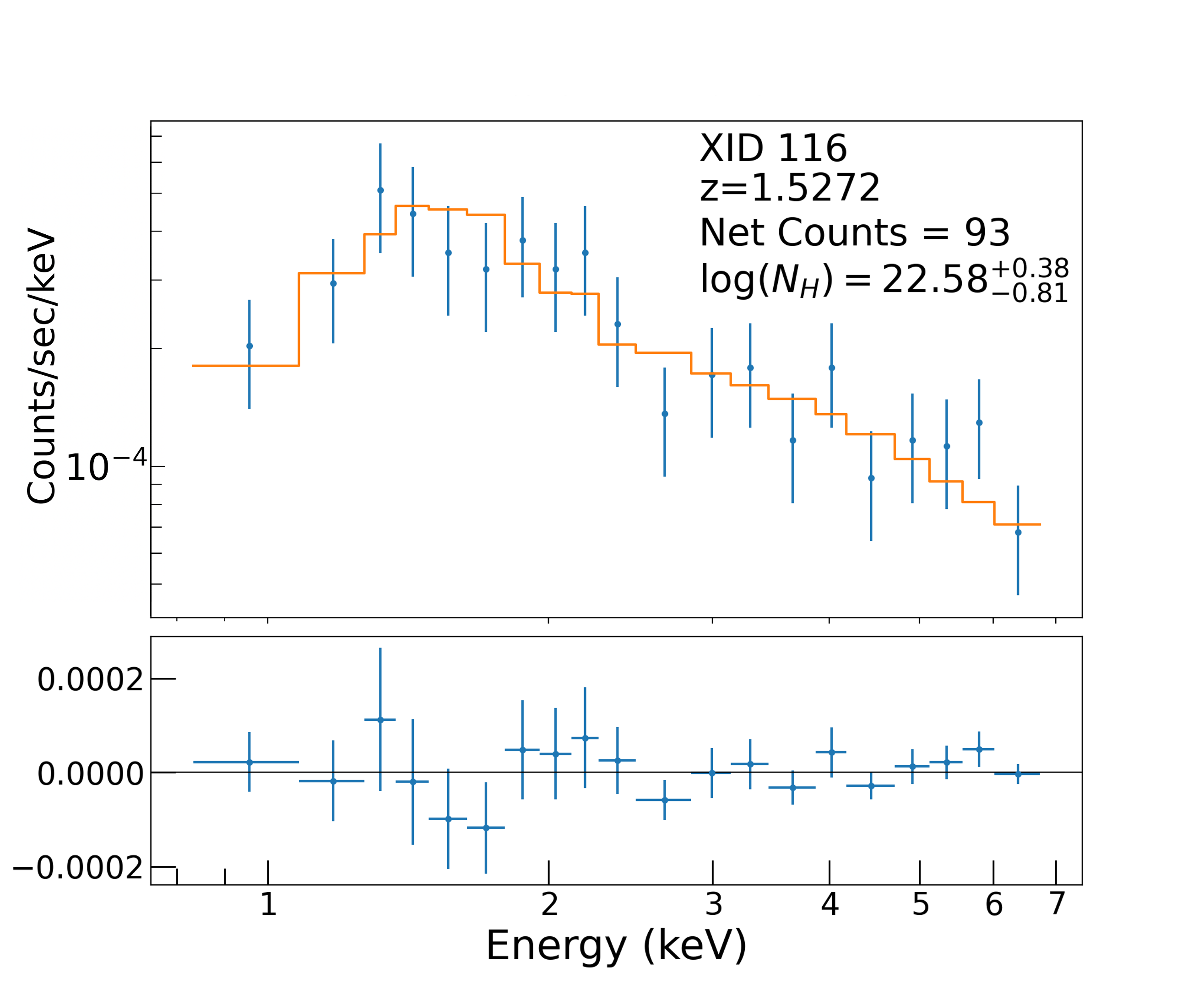}
\end{subfigure}
\begin{subfigure}{.49\textwidth}
  \centering
  \includegraphics[width=.98\linewidth]{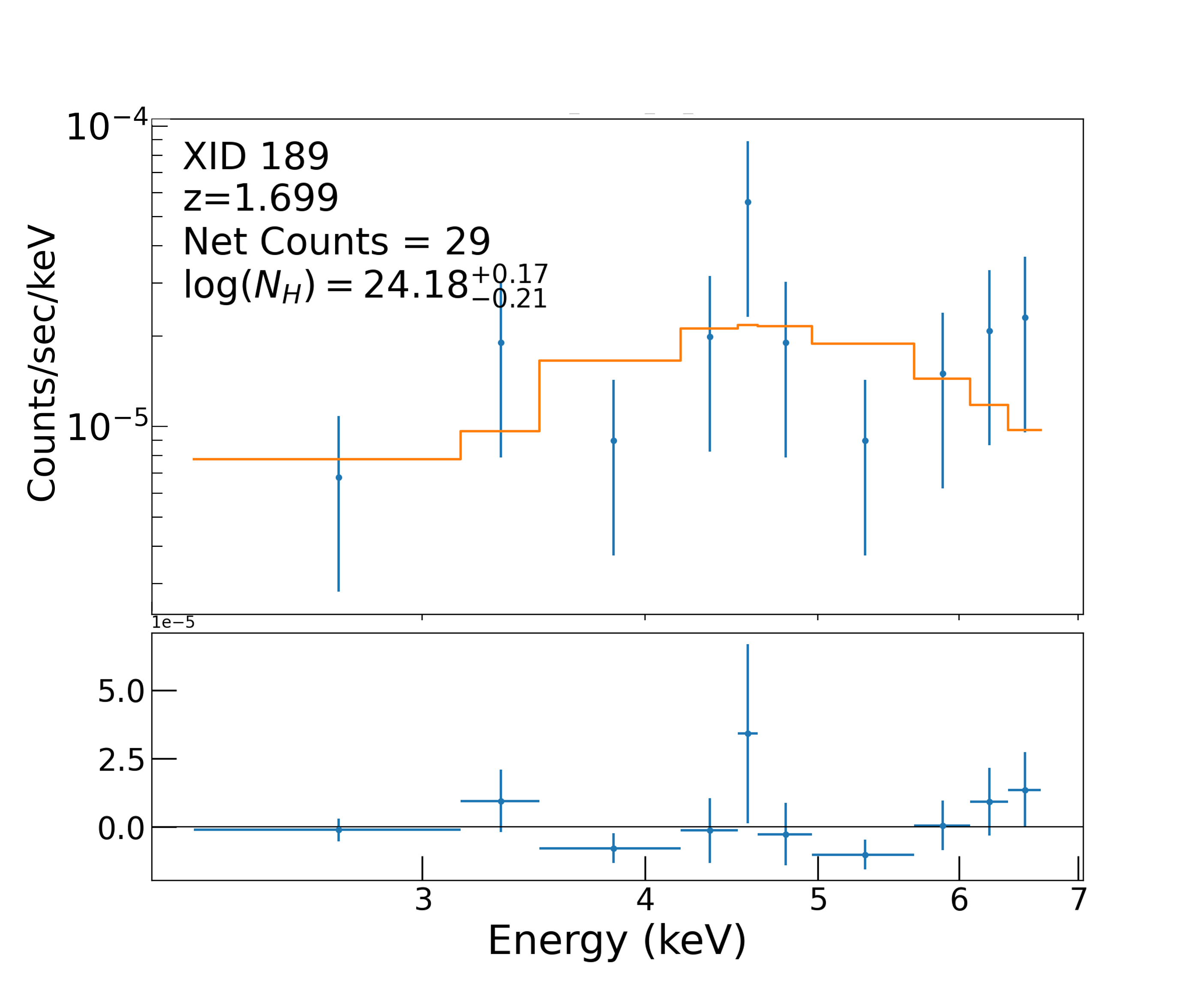}
\end{subfigure}
\caption{X-ray spectra in the 0.5-7 keV energy range (blue points) and best-fit models (orange solid lines) for four representative objects in the sample. In the lower panel, residuals are shown. The obscuration levels range from unobscured to heavily obscured. The lower panels show the residuals (data-model) of the fit.}
\label{fig:x_spectra_fit}
\end{figure*}

We first performed the fit procedure for the 39 objects with more than 150 counts. In Figure \ref{fig:gamma_hist}, we show the resulting photon index distribution; when fitted as a Gaussian, we found $\langle \Gamma \rangle=1.89$ and a standard deviation $\sigma_{\Gamma}=0.36$. So, we assume $\Gamma=1.9$ as a fixed parameter in fitting the objects with less than 150 counts. This value is also consistent with average values of the photon index $\Gamma$ found in the literature \citep{Mainieri07,Lanzuisi2013, Marchesi16,Liu_17}. The free parameters of the fit are then three (the power-law slope, the normalization, and the column density) when the objects have more than 150 counts, and two (the power-law normalization and the column density) otherwise. We derived 90$\%$ uncertainties on the fitted parameters with \textit{sherpa} $get\_conf()$. In Fig.~\ref{fig:x_spectra_fit} we show four representative X-ray spectra in our sample.
For 131 objects out of 243, the fit procedure returned only upper limits for the column density $N_{\rm H}$; for the others, we obtained $N_{\rm H}$  estimates with upper and lower bounds.
We note that an absorbed power-law model may not be an accurate representation of the X-ray spectra of the most heavily obscured, Compton-thick AGNs ($N_{\rm H}>10^{24}$\cm), where reflection components may dominate over the transmitted ones \citep{Comastri10, Marchesi18}. Nonetheless, the primary objective of this study is to determine the fraction of obscured AGNs using absorption thresholds of 10$^{22}$ or 10$^{23}$cm$^{-2}$. In this regard, we contend that an absorbed power-law model is well-suited for discerning whether the obscuration of an object exceeds the aforementioned thresholds.

The Fe $K \alpha$ line at 6.4 keV is a common feature of X-ray AGN spectra; the more obscured an object is, the more prominent this feature becomes, given the suppression of the primary continuum. Therefore, we expect to find it in a fraction of objects. To check for this presence, we performed the spectral fitting again, adding a new component to the source model, to search for the presence of such line at 6.4 keV. We apply two different strategies depending on whether the source has a spectroscopic redshift determination or a photometric one. %For the first kind of object, we know that the redshift estimate is well-constrained. Therefore, the only free parameter is the line normalization.
\begin{table}[ht]
%	\caption{Fe K$\alpha$ line - fixed$z$- width = 0.05 keV} % title of Table
	\centering % used for centering table
	\setlength{\tabcolsep}{2.5 pt}
	\begin{tabular}{c c c c c c} % centered columns (7 columns)
		\hline\hline %inserts double horizontal lines
		XID &$z$ & CNTS & $N_{\rm H}$   & $\Gamma$  & EW (keV) \\ [0.5ex] % inserts table
		%heading
		\hline % inserts single horizontal line
		2 & 0.628 & 827 & $1.6^{+0.6}_{-0.6}$ &  $1.9^{+0.2}_{-0.2}$  & $0.14^{+0.09}_{-0.06}$ \\
		
        4 & 2.013 & 259 & $< 4$ & $1.9^{+0.4}_{-0.3} $ & $0.19^{+0.12}_{-0.10}$\\
        
        8 & 2.78 & 102 & $31.53_{-14.82}^{+15.91}$ & 1.9 & $0.21_{-0.08}^{+0.12}$ \\
        
        31 & 2.377 & 154 & $<2.7$ & $2.0^{+0.3}_{-0.3} $ & $0.20^{+0.17}_{-0.11}$\\
        
		44 & 1.486 & 195 & $3.2^{+3.2}_{-2.9}$ &  $2.4^{+0.5}_{-0.5}$ & $0.24^{+0.42}_{-0.12}$\\
		
        70 & 0.764 & 228 & $4.1^{+2.4}_{-2.2}$ & $1.7^{+0.5}_{-0.5}$ &
        $0.21^{+0.21}_{-0.12}$\\
        
        73 & 2.171 & 226 & $<4.3$ & $2.0^{+0.4}_{-0.3}$ & $0.24^{+0.12}_{-0.10}$\\
        
        114 & 0.533 & 159 & $< 0.9$ & $2.2^{+0.6}_{-0.4}$ & $0.74^{+0.50}_{-0.42}$\\
        
        115 & 0.76 & 76 & $7.7^{+4.0}_{-3.0}$ & $1.9$ & $0.41^{+0.24}_{-0.17}$\\ [1ex] % [1ex] adds vertical space
		\hline %inserts single line
	\end{tabular}
   
\caption{Best fit parameters for the 9 objects with spectroscopic redshifts where a significant Fe K$\alpha$ line is detected at 6.4 keV. \\ The counts are the net full counts; the photon index $\Gamma$ is free to vary if the net counts of the source are more than 150, while it is fixed if they are less then 150. The column density $N_{\rm H}$  is shown in units of $10^{22} $cm$^{-2}$. The equivalent width of the emission line is shown in keV. Errors are at the 90\% confidence level. \label{Tab:table_fe_fixedE_width005} }	
\end{table}
%For the second kind, the redshift estimate has much larger uncertainties. Therefore we allow the redshift to vary, so it becomes an additional free parameter, and we link it to the power-law component redshift, imposing for them to have the same value. We chose to fix the width of the emission line to 0.05 keV, thus assuming no significant line broadening. \\
For objects with spectroscopic redshifts, we perform a new fit with the same model as before but with the addition of a single Gaussian line with 0.05 keV width. We considered the presence of the line to be significant when compared to the statistic of the best fitting simple absorbed power-law model, we obtain $\Delta C >2.7$, as we are adding one free parameter to the fit, the line normalization. This corresponds to 90$\%$ confidence level for 1 parameter of interest \citep[see, e.g.][] {Avni76, Tozzi06, Brightman14}. This happens for 9 objects out of 135: XID 2, XID 4, XID 8, XID 31, XID 44, XID 70, XID 73, XID 114, XID 115. For these objects, we also derived the rest-frame equivalent width of the Fe K$\alpha$ line. The results are shown in Table \ref{Tab:table_fe_fixedE_width005}.

\begin{table}[ht]
	%\caption{Fe K$\alpha$ line - free E - width = 0.05 keV} % title of Table
	\centering % used for centering table
	\setlength{\tabcolsep}{3pt}
	\begin{tabular}{c          c c c c c         c} 
		\hline\hline %inserts double horizontal lines
		XID &   $z_{phot}$  & CNTS & $N_{\rm H}$  & $\Gamma$ &  EW & $z_{line}$\\ [0.5ex] % inserts table
		%heading
		\hline % inserts single horizontal line

        46 & $1.64_{-0.44}^{+1.08}$ & 148 & $3.0_{-2.7}^{+3.1}$ & 1.9 & $0.38_{-0.17}^{+0.22}$ & $ 1.45_{-0.05}^{+0.05}$\\
        
        137 & $1.98_{-0.48}^{+3.2}$ & 55 & $10.3_{-8.3}^{+11.3}$ & 1.9 & $0.6_{-0.2}^{+0.3}$ & $ 2.23_{-0.07}^{+0.08}$\\
        
        167 & $2.94_{-0.72}^{+0.72}$ & 22 & $< 237$ & 1.9 & $0.92_{-0.50}^{+2.63}$ & $3.32_{-0.12}^{+0.13}$\\
        
        193 & $0.61_{-0.15}^{+0.13}$ & 35 & $5.5_{-2.7}^{+4.0}$ & 1.9 &  $0.54^{+0.44}_{-0.33}$ & $0.68_{-0.06}^{+0.04}$\\
        
        200 & $1.49_{-0.21}^{+1.57}$ & 112 & $< 2.1 $ & 1.9 & $0.3_{-0.1}^{+0.1}$ & $1.43_{-0.06}^{+0.07}$\\
        
        205 & $2.34_{-1.78}^{+2.44}$ & 38 & $104_{-61}^{+254}$ & 1.9 &  $12.8_{-7.4}^{+12.2}$ & $2.83_{-0.10}^{+0.15}$\\
        
        345 & $0.44_{-0.01}^{+0.18}$ & 180 & $< 9$ & $0.9_{-0.9}^{+2.1}$ &  $7.2_{-5.7}^{+14.0}$ & $0.62_{-0.02}^{+0.03}$\\ [1ex] % [1ex] adds vertical space
		\hline %inserts single line
	\end{tabular}
\caption{Same as Table 1, with the addition of the best-fitting line redshift ($z_{line}$) and associated 90\% uncertainties. \label{table_fe_freedE_width005} 
}
\end{table}

For objects for which we only have a photometric redshift estimate, the uncertainties on the redshift value are much bigger.
Therefore, in searching for a significant Fe K$\alpha$ line, we let the redshift of the model be a free parameter. We performed the fit with a single power-law model with the addition of a single Gaussian line with a fixed 6.4\,keV energy and a fixed 0.05 keV width, imposing the line redshift to be the same as the absorbed power-law. In this case, there are two additional parameters to the fit, which are the redshift and the line normalization. Therefore, we consider the presence of the emission line significant if the difference in the statistic is $\Delta C >5.4$. We found this to be true for 7 objects: XID 46, XID 137, XID 167, XID 193, XID 200, XID 205, and XID 345, whose properties are shown in Table \ref{table_fe_freedE_width005}. From this fit, we derive a redshift estimate, which in all cases is consistent with the photometric one, but provides a much smaller uncertainty. The average uncertainty on the redshift estimate for these objects goes from 0.94 in the photometric case to 0.07.  We note that for the object XID 205, which has a photometric redshift estimate of $z = 2.34^{+2.44}_{-1.78}$, we get an X-ray redshift estimate of $z = 2.82^{+0.15}_{-0.11}$, which is consistent with the redshift of the large-scale structure discovered in the field, $z=2.78$ \citep{Marchesi23}. 

Often a double power-law component is needed to fit AGN X-ray spectra, to model scattered emission which is typically found in obscured sources \citep{Ueda07}.
%Often a soft excess, which can have multiple physical origins, is detected in AGN X-ray spectra. 
To test for the presence of this component, we assumed a phenomenological model and we performed again the fit adding a power-law component with the same photon index as the main one, with no absorption and with a multiplicative constant whose maximum value was fixed at 0.3. Therefore, we only have one additional parameter, the multiplicative constant. We looked for objects for which $\Delta C > 2.7$ but we found none. This differs from the results in previous studies, where at least a few percent of objects are usually found to have a significant double power-law component \citep[e.g., ][]{Marchesi16}.

This might be caused by the decrease in the effective area of the Chandra telescope at energies below $\sim$1.5\,keV, mostly caused by the deposition of materials on the \textit{Advanced CCD Imager Spectrometer} (ACIS) detector.

\begin{figure*}
\centering
\begin{subfigure}{.49\textwidth}
  \centering
  \includegraphics[width=.99\linewidth]{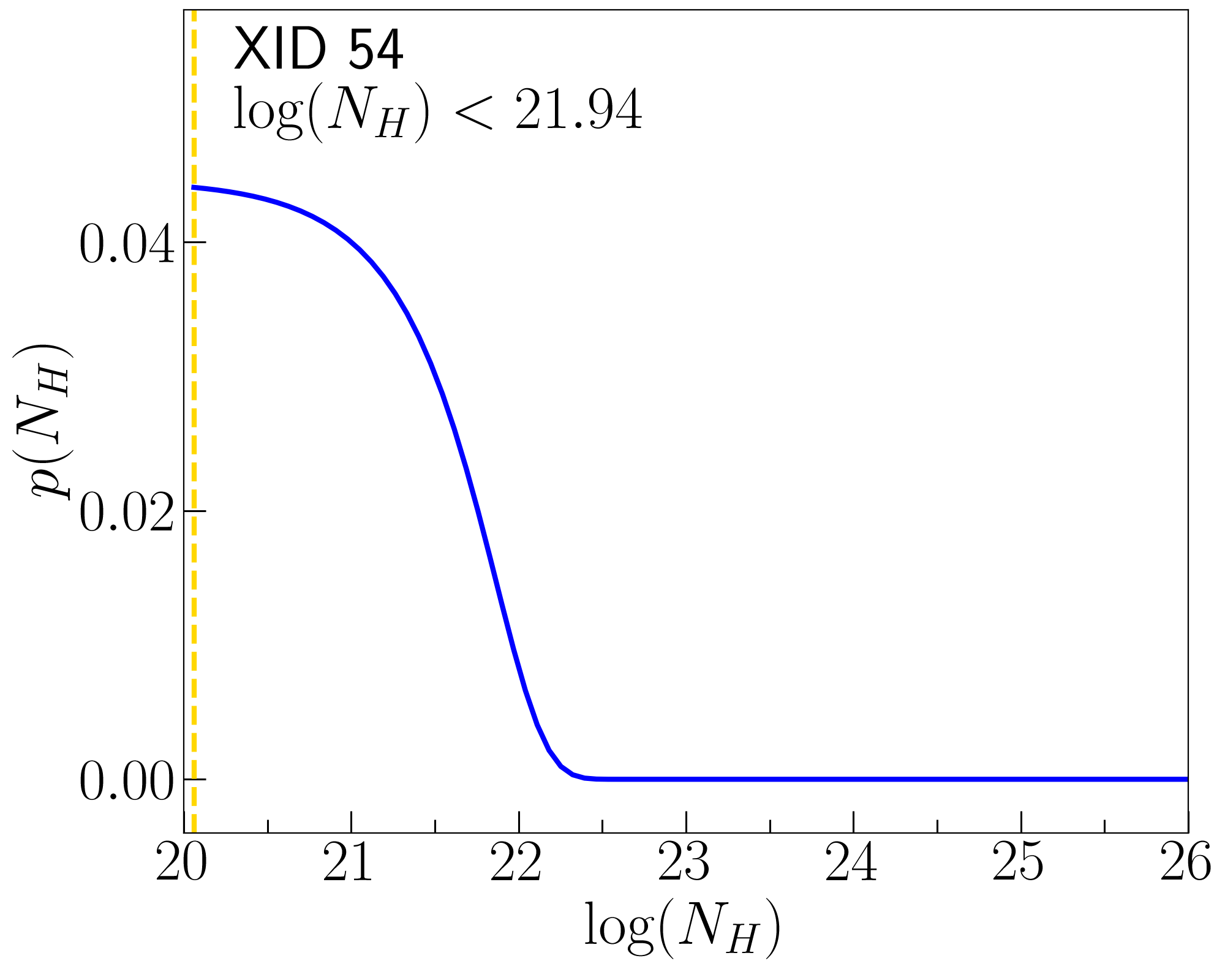}
  %\caption{$N_{\rm H}$  probability distribution for objects XID 54. The best fit\\ is $N_H < 8.6\cdot10^{21}$cm$^{-2}$}
\end{subfigure}%
\begin{subfigure}{.49\textwidth}
  \centering
  \includegraphics[width=.99\linewidth]{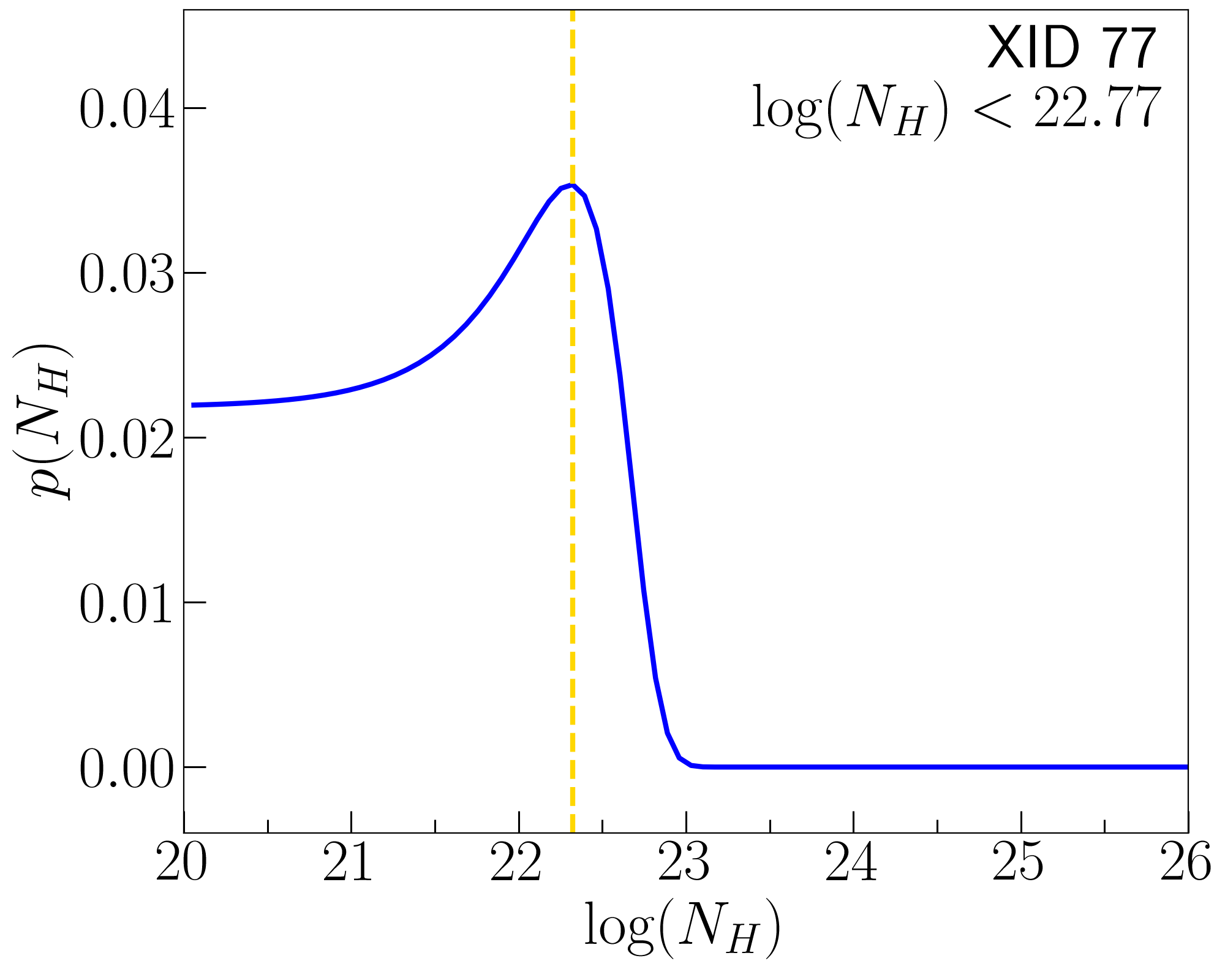}
  %\caption{$N_{\rm H}$  probability distribution for objects XID 77. The best fit\\ is $N_H < 3.8\cdot10^{22}$cm$^{-2}$}
\end{subfigure}
\begin{subfigure}{.49\textwidth}
  \centering
  \includegraphics[width=.99\linewidth]{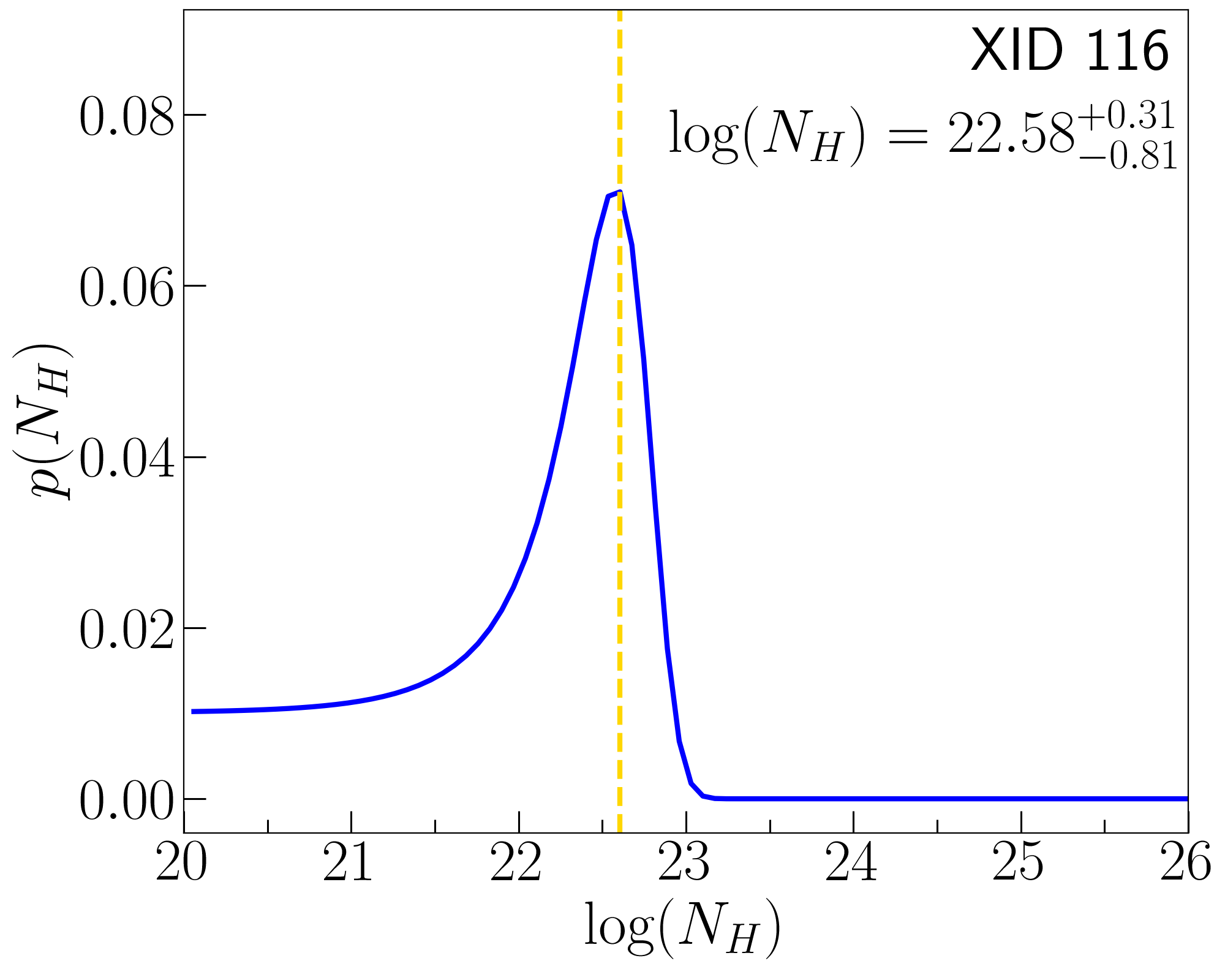}
  %\caption{$N_{\rm H}$  probability distribution for objects XID 116. The best fit \\is $N_H < 4.3\cdot10^{22}$cm$^{-2}$}
\end{subfigure}
\begin{subfigure}{.49\textwidth}
  \centering
  \includegraphics[width=.98\linewidth]{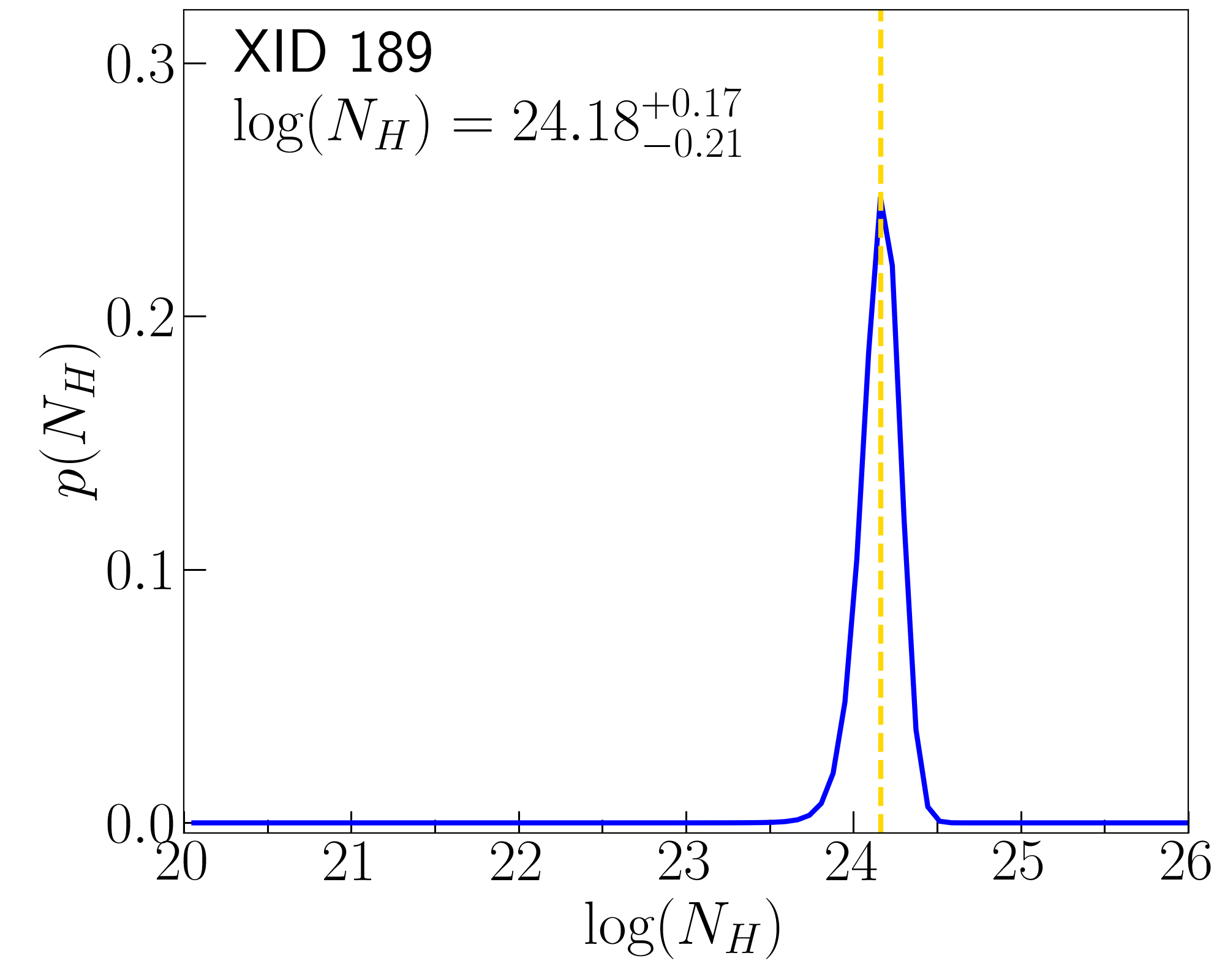}
  %\caption{$N_{\rm H}$  probability distribution for objects XID 189. The best fit \\is $N_H = 1.5^{+0.7}_{-0.6}\cdot10^{24}$cm$^{-2}$}
\end{subfigure}
\caption{$N_{\rm H}$ probability distributions for the four objects shown in Fig \ref{fig:x_spectra_fit}. The yellow dashed lines show the values of NH at which the minimum of the fit statistic is found. For the two objects in the upper panels, we could only derive upper limits to the  $N_{\rm H}$ measurements, whereas for the two objects in the lower panels, a significant ($>$90\% c.l.) column density was measured.}
\label{fig:pnh_distr}
\end{figure*}

\subsection{Column density probability distributions}\label{sec:nh_distr}
From the spectral fit, we obtain for each object a $N_{\rm H}$  estimate; for 131 out of 243 objects, the estimate is an upper limit for the $N_{\rm H}$ value, while for the others we have a best fit $N_{\rm H}$ value with upper and lower bounds. We can better understand the $N_{\rm H}$  estimates by deriving the $N_{\rm H}$  probability distributions for the objects in our sample. To do so, we used the \textit{sherpa} command $int\_proj$ to compute the fit statistic C as the $N_{\rm H}$  parameter is varied from $10^{19}$ to $10^{26}$cm$^{-2}$, using a logarithmic step of $\Delta \log(N_H)$ = 0.07. Given the statistic values, we derived the probability distribution $p(\log(N_{\rm H})\propto exp(-C/2))$ and normalized its integral to one. In Figure \ref{fig:pnh_distr} we show, as an example, the $N_{\rm H}$  probability distributions of the objects shown in Fig. \ref{fig:x_spectra_fit}. For XID 54 and XID 77, the fitting procedure returns an upper limit for the $N_{\rm H}$  estimate. However, one can see that the probability distributions are very different: for XID 54, each $N_{\rm H}$  value below $\sim10^{22}$cm$^{-2}$ is more or less equally likely; for XID 77 and XID 116, instead, there is a clear peak of the probability distribution around $\sim 3\times10^{22}$cm$^{-2}$, although the fit was not able to retrieve a lower bound to the $N_{\rm H}$  estimate. This is true for more than half of the objects in the sample: the $N_{\rm H}$  probability distributions are in many cases asymmetric, with low $N_{\rm H}$  values having a higher probability even when the peak of the distribution is at high $N_{\rm H}$  values. XID 189\footnote{Which, we note, is also the central object of the protocluster described in \cite{Gilli19}}, instead, presents a case in which the best-fitting $N_{\rm H}$ is well constrained, with a 90\% lower limit higher than zero. The fit, in this case, is indeed able to retrieve both an upper and a lower bound for $N_{\rm H}$. 
As can be seen with these examples, the probability distribution is a more accurate way to describe the column density of a source, compared with the nominal $N_{\rm H}$ value that we obtain from the fit. We, therefore, chose to use the $p(\log(N_{\rm H}))$ to derive the obscured fractions, as we will discuss in Section \ref{sec:obscured_fraction}. 
The $N_{\rm H}$  probability distributions for all the objects in the sample are available on the project webpage\footnote{\url{http://j1030-field.oas.inaf.it/xray_redshift_J1030.html}}.

\subsection{Results}\label{sec:spectral_results}
\begin{figure*}
\centering
\includegraphics[width=0.75\linewidth]{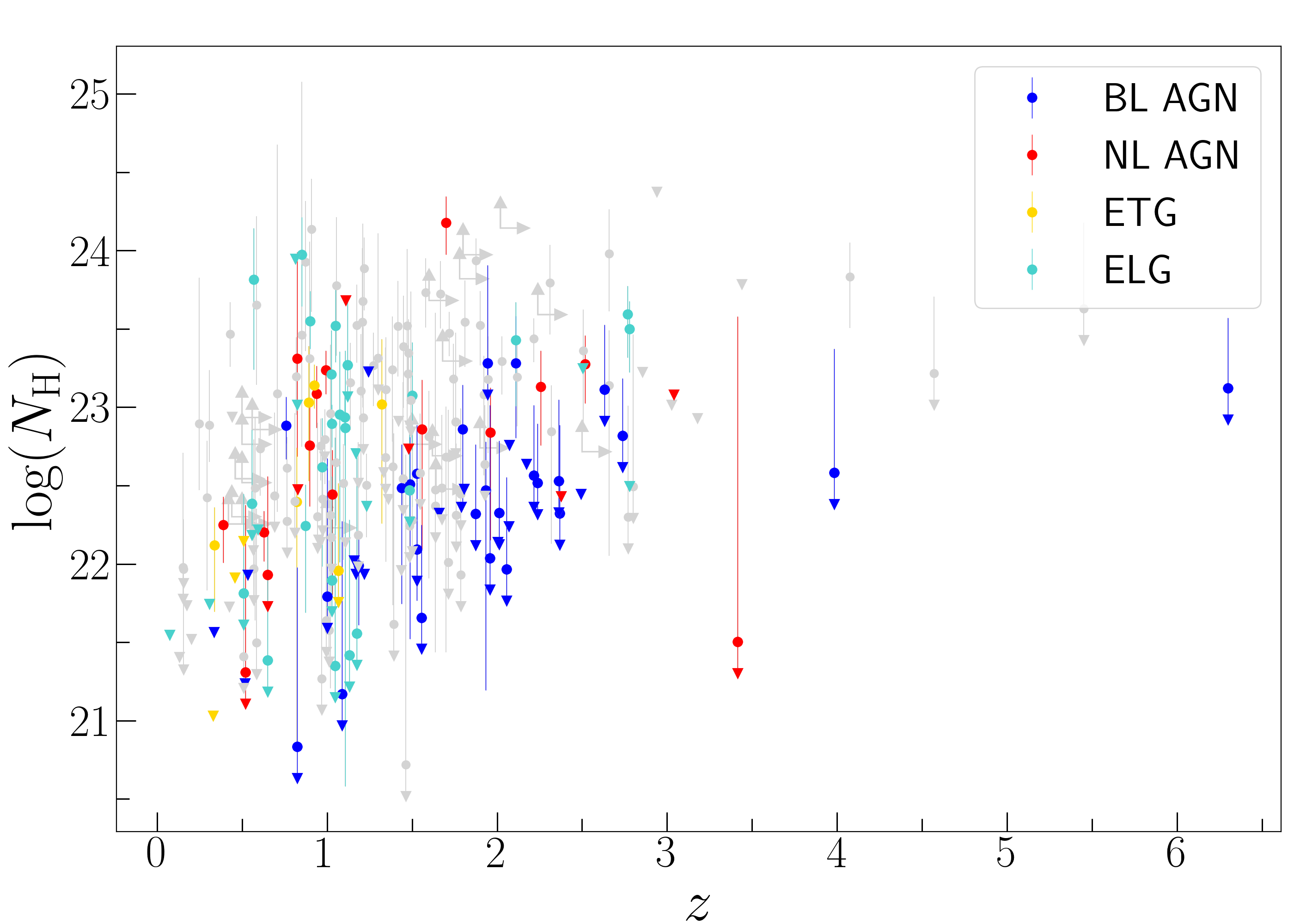}
\caption{$N_{\rm H}$  - redshift distribution of all the objects in the catalog. Up-right pointing arrows show the redshift and $N_{\rm H}$ lower limits for objects with a flat photometric redshift probability curve. Upper limits are shown as down-pointing triangles. The color code stands for  spectral type: red: NL-AGN, blue: BL-AGN, yellow: Early Type Galaxies, aquamarine: Emission Line Galaxies, grey: no spectral identification}
\label{fig:NH_z_all}
\end{figure*}

At the end of our spectral analysis, we have derived the column density $N_{\rm H}$  for the 243 AGNs in the J1030 \textit{Chandra} field.
The catalog with the basic physical properties derived from our analysis is available online \footnote{ \url{http://j1030-field.oas.inaf.it/chandra_1030}}; in Table \ref{table_catalog} we show a portion of it. For each object, we provide the column density, the photon index, the (de-absorbed), rest-frame 2-10 keV luminosity, and relative 90\% uncertainties. In Figure \ref{fig:NH_z_all} we show the global $N_{\rm H}$  - redshift distribution for the sample, with objects shown with different symbols and colors depending on their spectral identification \citep{Marchesi_21}. A trend of $N_{\rm H}$  with redshift can be seen, with objects at higher redshift having on average higher $N_{\rm H}$ values. This is partly due to a selection effect, since moving towards higher redshifts, the photoelectric absorption cutoff moves outside the limit of the observing band, and it is, therefore, more difficult to constrain lower $N_{\rm H}$ values \citep{Civano05, Lanzuisi2013}. A thorough analysis of the obscured fraction trend with redshift that takes this factor into account is provided in the next Section. 

\begin{figure*}
\centering
\includegraphics[width=0.75\linewidth]{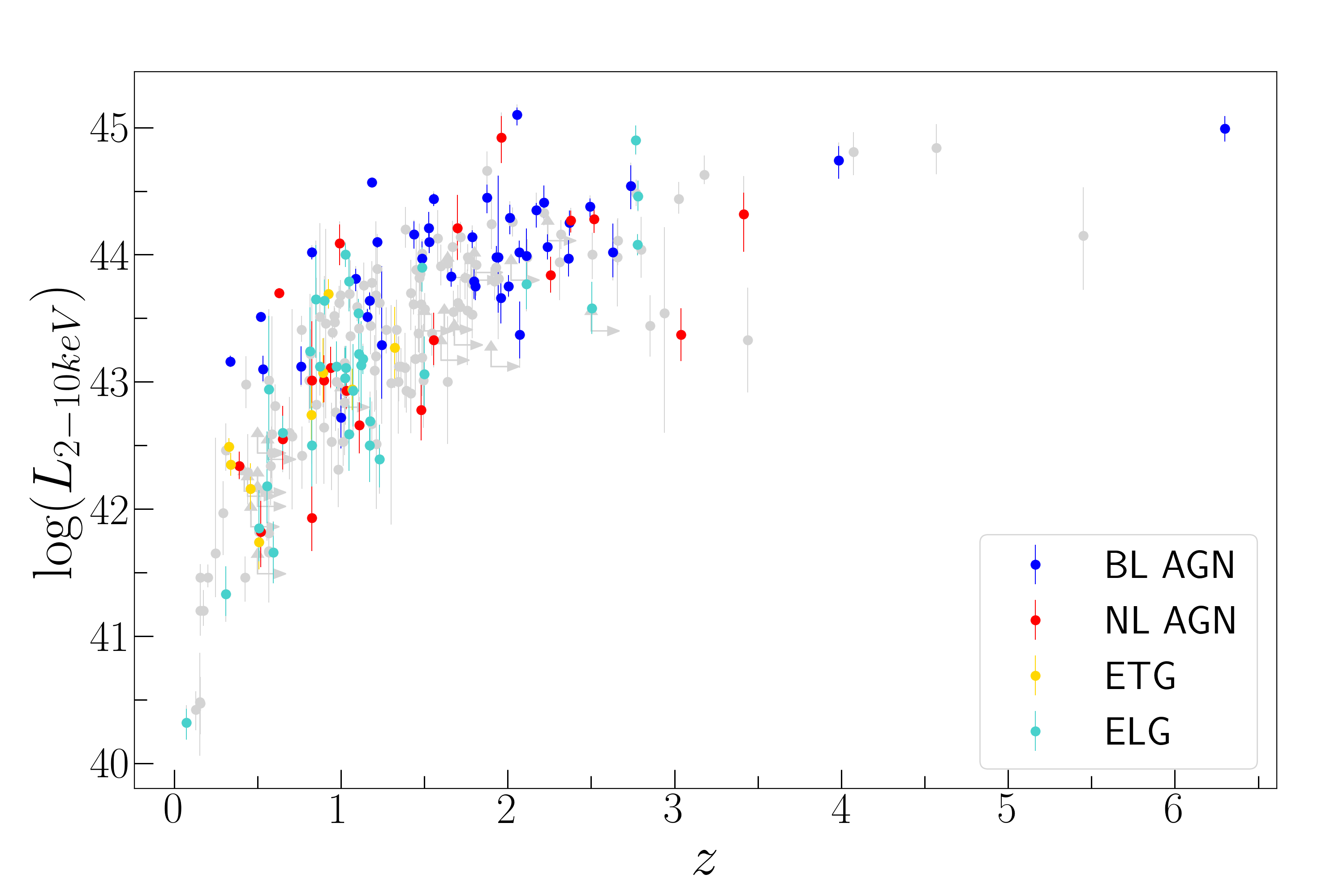}
\caption{Distribution of the intrinsic (de-absorbed), rest frame, 2-10 keV luminosity for the 243 objects in the catalog. Up-right pointing arrows show the redshift and luminosity lower limits for objects with a flat photometric redshift probability curve. The different colors code identify the spectral type and are the same as in Figure \ref{fig:NH_z_all}.}
\label{fig:Lx_z}
\end{figure*}

We note that the column densities that we obtain for objects for which we have a classification from the optical spectrum are consistent with the optical classifications themselves: Broad Line AGNs (in blue) have low column densities, and for $90\%$ of them  the spectral fit can only obtain an upper limit for $N_{\rm H}$, while Narrow-Line AGNs (in red) have higher average column densities and the fraction for which we get upper limit for $N_{\rm H}$ is $40 \%$. This fraction is $51\%$ for ELGs and $52\%$ for ETGs. The sources for which we obtain the higher $N_{\rm H}$  values are more likely to be those without an optical spectral classification (in gray), which is consistent with them being obscured and therefore not easily observed in the UV/optical. In Figure \ref{fig:Lx_z} we show the intrinsic (i.e de-absorbed) rest-frame 2-10 keV luminosity vs redshift, with the same classification code.

\begin{table*}[ht]
%	\caption{} % title of Table
	\centering % used for centering table
	\setlength{\tabcolsep}{3 pt}
	\begin{tabular}{c c c c c c c} % centered columns (7 columns)
		\hline\hline %inserts double horizontal lines
		XID &$z$ & CNTS & $\log(N_H)$ & $\Gamma$  & $\log(L_{2-10 keV})$ & cstat/dof\\ [0.5ex] % inserts table
		%heading
		\hline % inserts single horizontal line
		 & & & cm$^{-2}$ &  & erg/s & \\
		\hline
		1 & 3.18$^{+1.3}_{-1.3}$ & 252$^{+17}_{-16}$ & $<22.93$ & 1.52$^{+0.33}_{-0.23}$ & 44.63$^{+0.15}_{-0.08}$ & 144.1/172 \\
		
        2 & 0.6279 & 827$^{+30}_{-29}$ & 22.20$^{+0.14}_{-0.19}$ & 1.93$^{+0.22}_{-0.22}$ & 43.70$^{+0.04}_{-0.04}$ & 288.8/296 \\
        
        3 & 1.0974 & 164$^{+14}_{-13}$ & 22.52$^{+0.25}_{-0.49}$ & 2.12$^{+0.53}_{-0.48}$ & 43.59$^{+0.13}_{-0.11}$ & 99.6/134  \\
        
        4 & 2.0133 & 259$^{+17}_{-16}$ & $<22.79$ & 1.85$^{+0.36}_{-0.32}$ & 42.29$^{+0.13}_{-0.10}$ & 127.5/171 \\
        
        5 & 0.9679 & 37$^{+7}_{-6}$ & $<22.40$  & 1.9 & 42.76$^{+0.18}_{-0.15}$ & 265.7/289  \\
        
        6 & 0.5181 & 993$^{+33}_{-32}$ & $<21.24$ & 1.903$^{+0.12}_{-0.10}$ & 43.51$^{+0.03}_{-0.03}$ & 59.79/72 \\

	%\hline %inserts single line
	\end{tabular}
\caption{\textit{Chandra} J1030 spectral catalog. For each object, we provide the redshift (which is derived from spectroscopy when provided without uncertainties, from photometry otherwise), the (0.5-7 keV) counts \citep[see][]{Nanni_18} the logarithm of the column density, the photon index (which is fixed to 1.9 when the counts are less than 150), the intrinsic rest-frame 2-10 keV luminosity and the value of the C-statistic over degrees of freedom. Six objects also have a redshift estimate derived from the presence of the Fe K$\alpha$ line (see Table \ref{table_fe_freedE_width005}). This table is available in its entirety in machine-readable form on the website. \label{table_catalog}}

\end{table*}

\section{Obscured fraction}\label{sec:obscured_fraction}
\begin{figure*}
\centering
\includegraphics[width=0.75\linewidth]{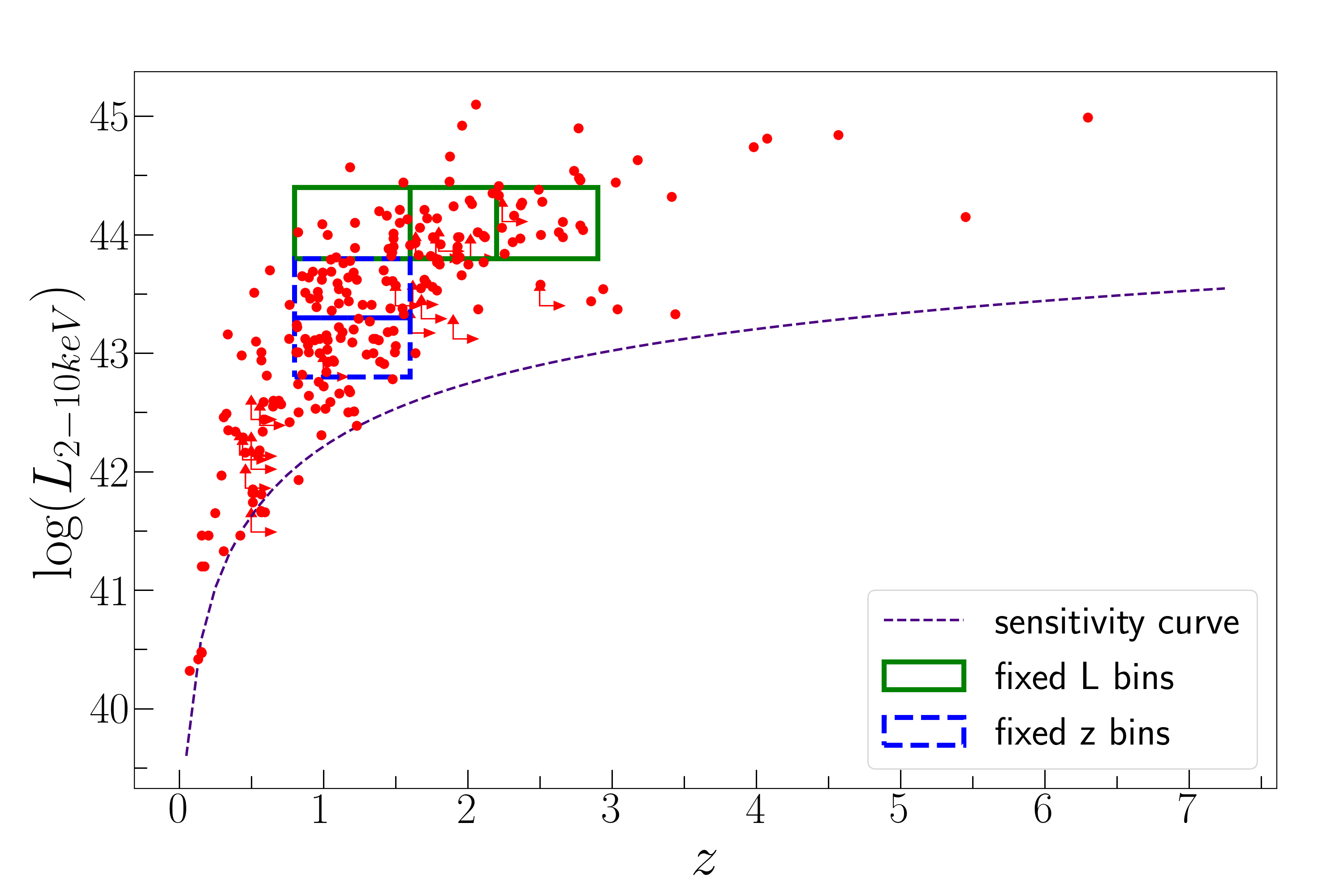}
\caption{Intrinsic rest-frame 2-10 keV luminosity as a function of redshift for the 203 Chandra J1030 sources detected in the 2-7 keV band. Up-right pointing arrows show the redshift and luminosity lower limits for objects with a flat photometric redshift probability curve.  In green, the subsample used for the analysis of the $N_{\rm H}$ -redshift evolution (Section~\ref{sec:obsc_frac_vs_z}); in blue, the subsample used for the analysis of the $N_{\rm H}$ -luminosity evolution (Section~\ref{sec:obsc_frac_vs_lx}). The dashed purple line represents the survey sensitivity curve, at 50\% of the field coverage \citep{Nanni_20}}.
\label{L_z}
\end{figure*}

Our goal here is to investigate the dependence of the column density $N_{\rm H}$ on redshift and luminosity. We have to consider that at different redshifts we sample
different average luminosities. In the literature, there is evidence of the obscuring fraction being a function of both redshift and luminosity \citep[see, e.g.,][]{Ueda14,Aird15,Ananna19}. Therefore, we need to perform our analysis at a fixed luminosity to derive the evolution of $N_{\rm H}$  with redshift, and at a fixed redshift to derive the $N_{\rm H}$ dependence on luminosity. We considered the intrinsic luminosity-redshift plane, which can be seen in Figure \ref{L_z}, and we selected only the objects which have a hard band detection, which is 203 out of 243, to get a uniform selection function and apply reliable correction to go from observed to intrinsic obscured fractions (see Section \ref{sec:obsc_frac_vs_lx}). 

For the luminosity-dependence analysis, we selected the subsample shown in blue, where the average redshift is $\sim 1.2$ in each bin. This subsample can be divided into three luminosity bins, with $42.8 < \log(L_{2-10 keV}) < 43.3$, $43.3 < \log(L_{2-10 keV}) < 43.8$, and $43.8 < \log(L_{2-10 keV}) < 44.5$, respectively. In each bin, we have 38, 32, and 18 objects, respectively. Out of these objects, the ones with a spectroscopic redshift estimate are 19 out of 38 in the first, 11 out of 32 in the second, and 11 out of 18 in the third bin.\\
For the redshift dependence analysis, we selected a subsample of objects, shown in green, with an average luminosity of $10^{44}$ erg/s. This subsample is then divided into three subsamples with redshift $0.8 <z<1.6$, $1.6 < z < 2.2$, and $2.2 < z < 2.8$. In each redshift bin, the average luminosity is $\sim10^{44}$ erg$/$s, and we have 18, 24, and 20 objects per bin, respectively. Out of these objects, the ones with a spectroscopic redshift estimate are 11 out of 18 in the first, 11 out of 24 in the second, and 14 out of 20 in the third bin.\\
These bins were selected to maximize the source statistics while keeping the best completeness in each bin. We note that, for the few objects with a flat redshift probability distribution (4 out of a total of 60 objects in the five different bins), we used their best redshift estimate to determine whether they belong to a certain bin.

\subsection{Obscured fraction dependence on 2--10 keV luminosity}\label{sec:obsc_frac_vs_lx}
We want to derive the obscured fraction $f_{22}$, which is the fraction of objects with a column density $N_H > 10^{22}$cm$^{-2}$, and $f_{23}$, the fraction of objects with a column density $N_H > 10^{23}$cm$^{-2}$. For each object, we could simply use the best fit value of $N_{\rm H}$  as the $N_{\rm H}$  estimate. However, this does not take into account how \textit{likely} it is for the true $N_{\rm H}$  value for a given object to be that of the nominal result of the fit. Furthermore, for objects with similar $N_{\rm H}$ values (or upper limits), the probability distributions can vary significantly from one object to another, as shown before.

Considering all of this, we derived the obscured fractions using the probability distribution functions described in Section \ref{sec:analysis}. For each object, we considered the fraction of $p(\log(N_{\rm H}))$ at $N_{\rm H}$  values higher than $10^{22}$cm$^{-2}$ ($10^{23}$cm$^{-2}$). We summed all the fractions for the objects in a given luminosity bin and got an estimate of the number of obscured sources that correctly takes into account the probability distribution functions. By dividing this number by the total number of objects in the bin, we obtain the observed obscured fraction. We performed this for the two different obscuration thresholds ($10^{22}$cm$^{-2}$ and $10^{23}$cm$^{-2}$) and for each luminosity bin.

\begin{figure*}
\centering
\begin{subfigure}{.5\textwidth}
  \centering
  \includegraphics[width=.99\linewidth]{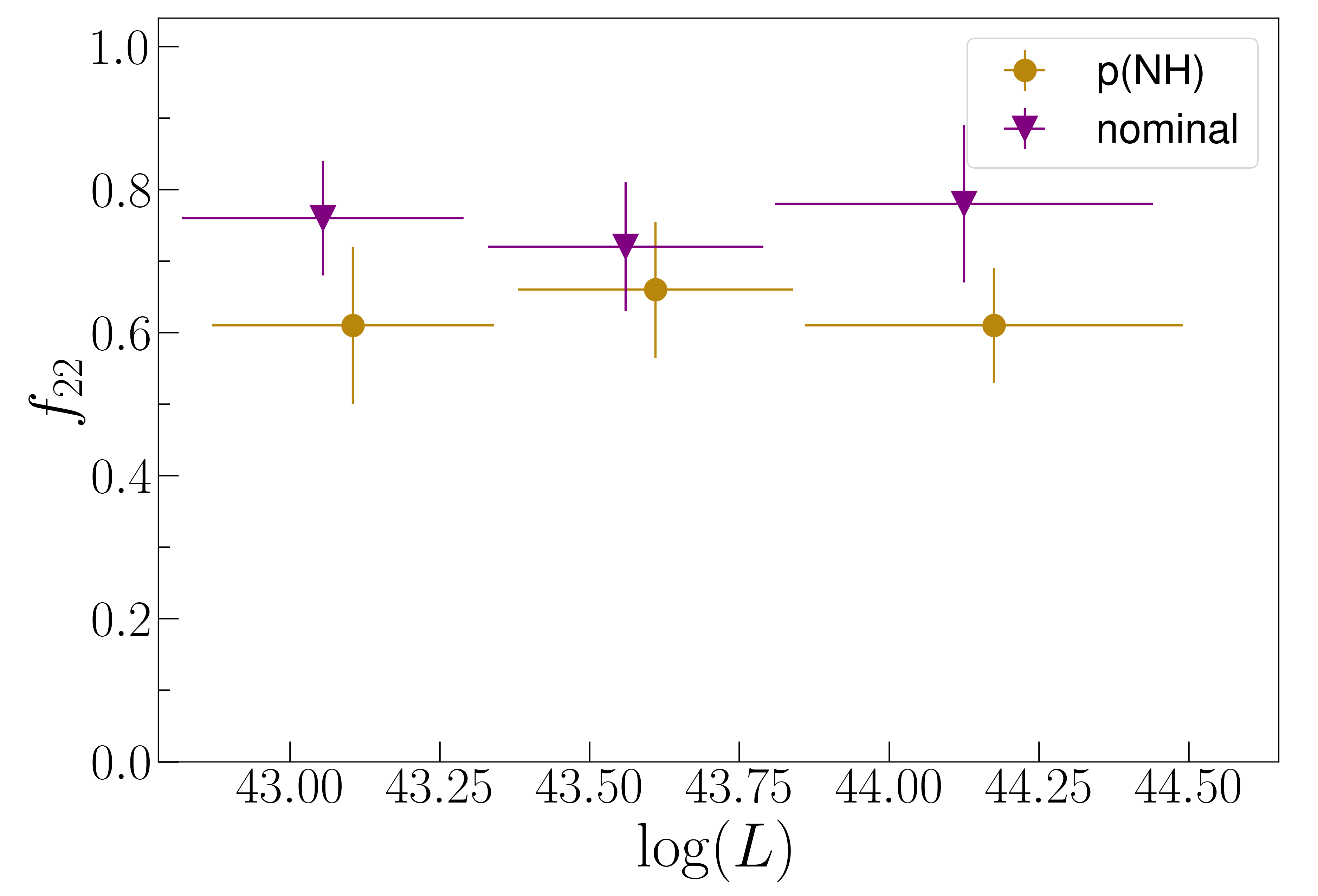}
  %\caption{Threshold: $10^{22} $cm$^{-2}$}
\end{subfigure}%
\begin{subfigure}{.5\textwidth}
  \centering
  \includegraphics[width=.99\linewidth]{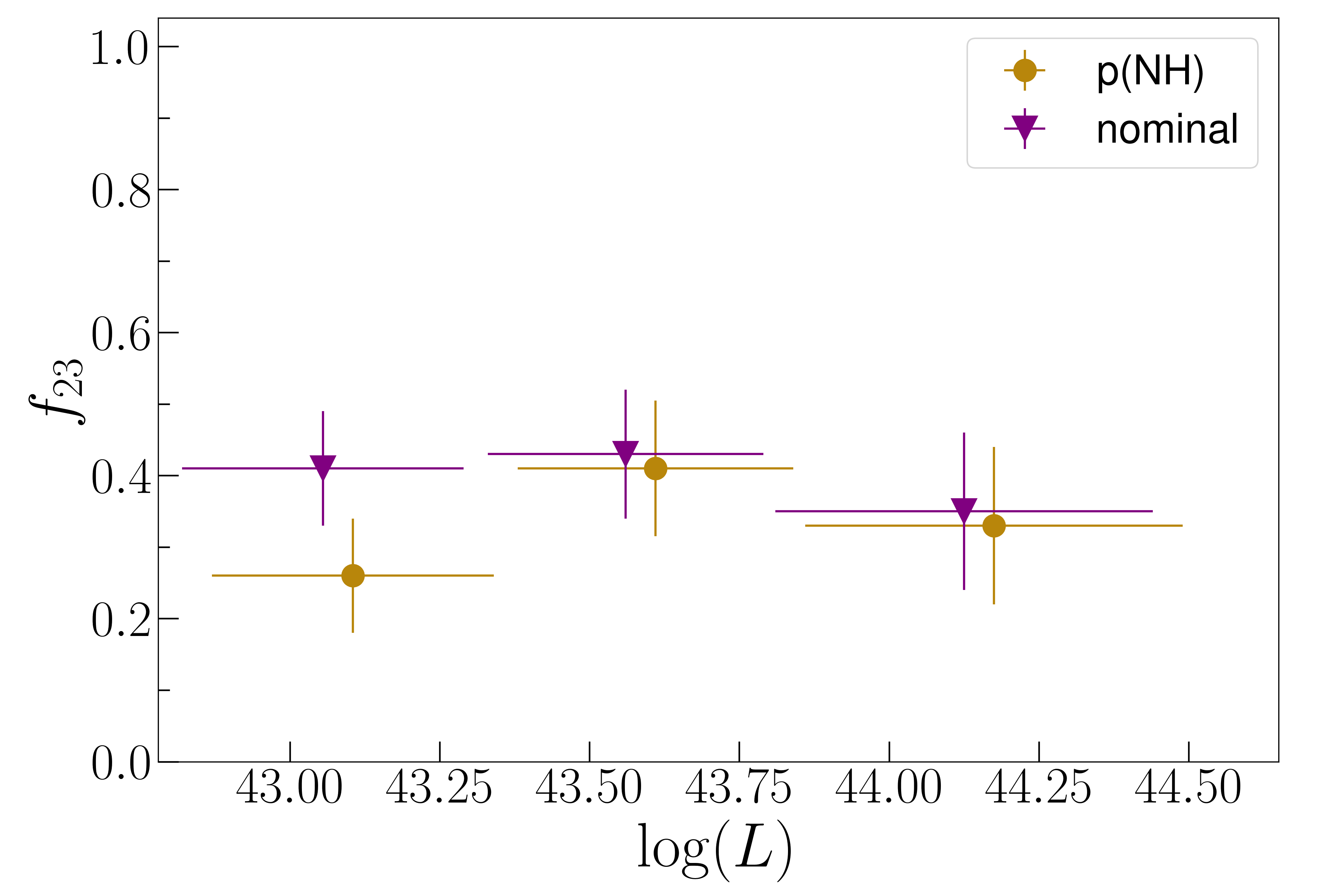}
 % \caption{Threshold: $10^{23} $cm$^{-2}$}
\end{subfigure}
\caption{Fraction of obscured  $z\sim 1.2$ AGN as a function of intrinsic 2-10 keV luminosity. Purple triangles show the observed obscured fractions derived using the nominal $N_{\rm H}$ value derived from the fit. Golden circles are the ones obtained using the probability distribution of $N_{\rm H}$. The second set of points is shifted by 0.05 on the log(L) axis for visual clarity. 1-$\sigma$ uncertainties, derived with the bootstrapping procedure, are shown. This Figure highlights the relevance of using the probability distributions in deriving the obscured fraction of AGNs. These results are still not corrected for the survey sky coverage (for those, see Figure \ref{fig:f_L_withothers}). \textit{Left}: obscured fraction derived using $N_{\rm H} > 10^{22}$\cm\ as the threshold ($f_{22}$). \textit{Right}: obscured fraction derived using $N_{\rm H} > 10^{23}$\cm\ as the threshold ($f_{23}$).}
\label{fig:nom_pnh}
\end{figure*}

In Figure \ref{fig:nom_pnh} we show the comparison between the results obtained using this procedure, which is using the $p(\log(N_{\rm H}))$, and using the nominal values of $N_{\rm H}$. It can be seen that using the $p(\log(N_{\rm H}))$ we get obscured fractions $f$ systematically lower than the others (even by $\Delta f \sim 0.18$). This is expected, as most $N_{\rm H}$  probability distributions are skewed towards lower $N_{\rm H}$  values. Therefore, there are objects for which the nominal $N_{\rm H}$ value can be higher than the obscuration threshold, but that does not overall contribute much to the obscured population in terms of its probability distribution. The asymmetry of the probability distributions mainly depends on the lack of information at soft X-ray energies. Because of this, it is often possible in the fitting procedure to get a high obscuration level excluded, but it is not possible to distinguish between a non-obscured and a mildly-obscured object.

\begin{table}[ht]
\captionof{table}{Number of objects, average redshift, and fraction of AGN with $\log(N_{\rm H})>22$ ($f_{22}$) and $\log(N_{\rm H})>23$ ($f_{23}$) in three luminosity bin and relative uncertainties} % title of Table
	\centering % used for centering table
	\setlength{\tabcolsep}{3pt}
	\begin{tabular}{p{3cm}| p{0.4cm} p{0.65cm} p{1.7cm} p{1.7cm} } 
		\hline\hline %inserts double horizontal lines
		Bin & N & $\bar{z}$ & $f_{22}$ & $f_{23}$ \\ [0.5ex] % inserts table
		%heading
		\hline % inserts single horizontal line
        $42.8 < \log(L) < 43.3$ & 38 & 1.15 & 0.80 $\pm$ 0.11 & 0.65 $\pm$ 0.11 \\
        $43.3 < \log(L) < 43.8$ & 32 & 1.16 & 0.80 $\pm$ 0.10 & 0.78 $\pm$ 0.10 \\
        $43.8 < \log(L) < 44.5$ & 18 & 1.35 & 0.78 $\pm$ 0.08 & 0.39 $\pm$ 0.08\\
    %     \\[1ex] % [1ex] adds vertical space
		\hline %inserts single line
	\end{tabular}
	\label{table_unc_L_f22} 
\\[1ex]	
%\caption*{}
\end{table}

Regarding the uncertainties on these obscured fractions, we know that confidence intervals on sample proportions are usually derived using the binomial distribution. We can, for example, use the Wilson score interval \citep{Wilson27} to derive confidence intervals, which will depend, in each bin, on the number of objects in the bin and on the obscured fraction derived using the probability distributions. When doing so for the three luminosity bins, we get lower uncertainties around $\sim$0.5 and upper uncertainties around $\sim$0.10. However, this method takes into account the uncertainties related to the finiteness of the sample only and does not consider that the $N_{\rm H}$ estimates are not exact. To deal with this, we derived the uncertainties with a bootstrap procedure: for each bin, we randomly extract, from the bin, with re-entry, a number of objects equal to the bin size. Then, for each object, we extract a value for $N_{\rm H}$ from its probability distribution. We then compute the obscured fraction as the number of objects with $N_{\rm H}>10^{22}$cm$^{-2}$ over the total. We repeat this 10000 times and we obtain a $f_{22}$ (or $f_{23}$) distribution, from which we extract the peak and the 16\% and 84\% quantiles as the values for $f_{22}$ (or $f_{23}$) and the corresponding uncertainties. In this way, both the finiteness of the bin and the uncertainties on each $N_{\rm H}$ estimate are taken into account.

We now must consider that our survey is flux-limited. This means that we are likely to miss preferentially obscured (i.e. fainter) objects rather than unobscured ones. Therefore, the obscured fractions that we derive are only lower limits to the intrinsic obscured fraction, and the true value will be higher. We need to correct the obtained values for the number of objects that we are not observing (the so-called Malmquist bias).
To do so, we proceeded in the following way for each luminosity bin and for each obscuration threshold ($10^{22}$ and $10^{23}$ cm$^{-2}$): we considered the \textit{intrinsic} number of obscured an unobscured sources in a given redshift and luminosity range ($N_O^{int}$ and $N_U^{int}$, respectively) expected in the population synthesis model of the cosmic X-ray background (XRB) of \cite{Gilli07}. To derive them, we used the online tool \footnote{\url{http://www.bo.astro.it/~gilli/counts.html}} to compute the surface density - or integral number counts, $N(>S)$ - above any given 2-10 keV flux limit S of both obscured and unobscured AGN. The expected, intrinsic number of obscured and unobscured AGN in J1030 $N_O^{int}$, $N_U^{int}$, were obtained by multiplying the source surface density at $f_{2-10 keV}=10^{-20}$ cgs (i.e. at $\approx$ zero flux) by the geometric area of J1030. From the integral number counts, we then obtained the differential source counts dN/dS and folded them with the sky coverage A(S) of the J1030 survey \citep{Nanni_20} as $\int dn/dS A(S) dS$. Because the sky coverage is given in the 2-7 keV flux range, we convert it to the 2-10 keV range by assuming a power-law spectrum with a photon index of 1.4, which is the average \textit{observed} index for the AGN population.\\
In this way, we obtain $N_{O}^{obs}$ and $N_{U}^{obs}$, which are the expected \textit{observed} number of obscured and unobscured objects.
We then derived the intrinsic and the observed ratios between the number of obscured objects and unobscured ones, $R_{int} = N_{O}^{int} / N_{U}^{int}$ and $R_{obs} = N_{O}^{obs} /N_{U}^{obs}$. As we lose more obscured objects than unobscured ones when in the presence of a flux limit, $R_{int}$ will always be larger than $R_{obs}$. We can now derive $p = R_{obs}/R_{int}$ as the corrective parameter that we need to implement to go from our \textit{observed} obscured fraction to the \textit{intrinsic one}. This number is always smaller than one. 

If we now define the observed obscured fraction(s) as $f_{22}=N[10^{22} - 10^{26}] / N[10^{20} - 10^{26}]$ and $f_{23}=N[10^{23} - 10^{26}] / N[10^{20} - 10^{26}]$, we can derive the corrected fractions as:
\begin{equation}
    f_{22}^{corrected} = \frac{f_{22}}{f_{22} (1-p)+p}
\end{equation}
and we can derive $f_{23}^{corrected}$ in the same way. 
\begin{figure*}
\centering
\begin{subfigure}{.5\textwidth}
  \centering
  \includegraphics[width=.99\linewidth]{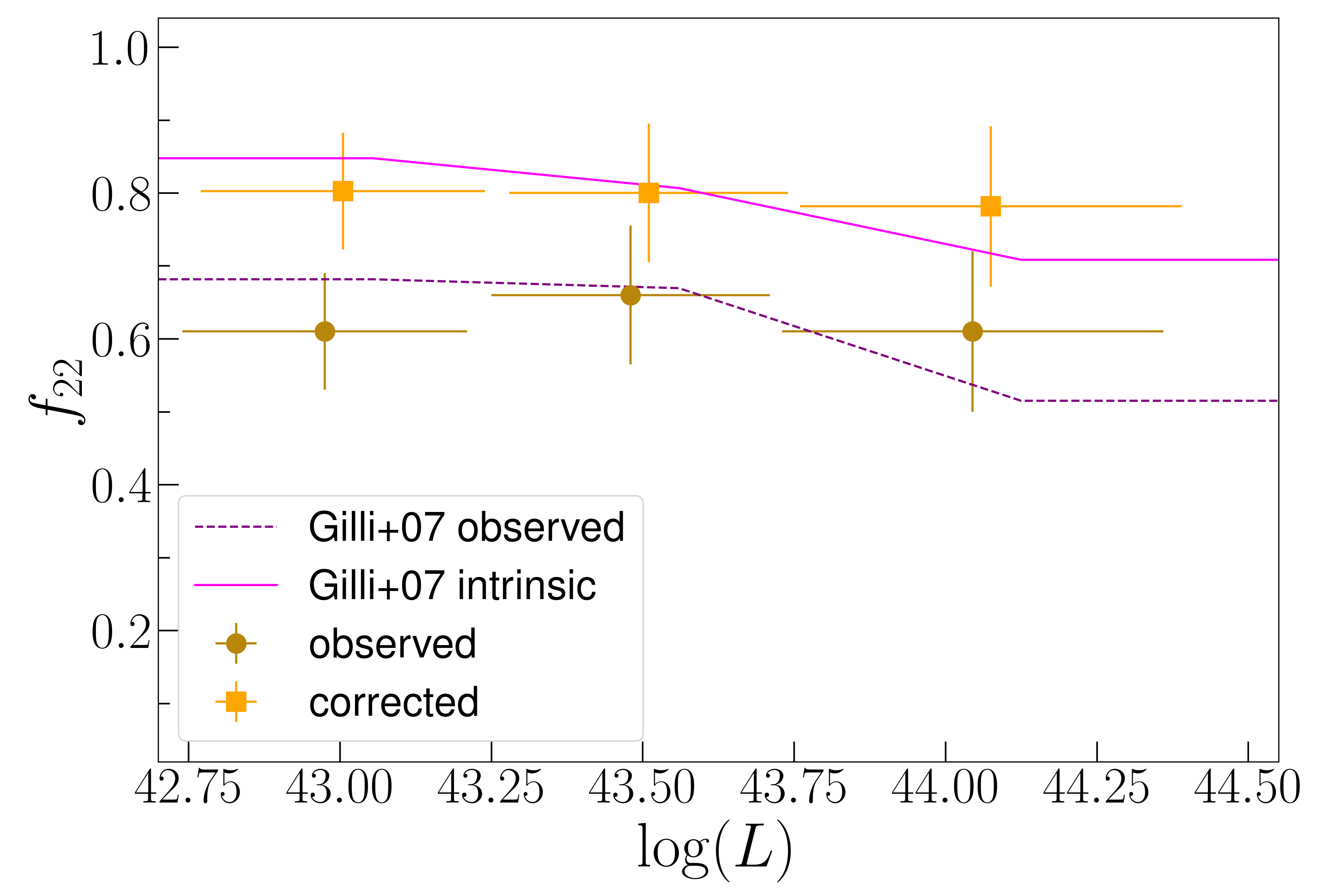}
  %\caption{Threshold: $10^{22} $cm$^{-2}$}
\end{subfigure}%
\begin{subfigure}{.5\textwidth}
  \centering
  \includegraphics[width=.99\linewidth]{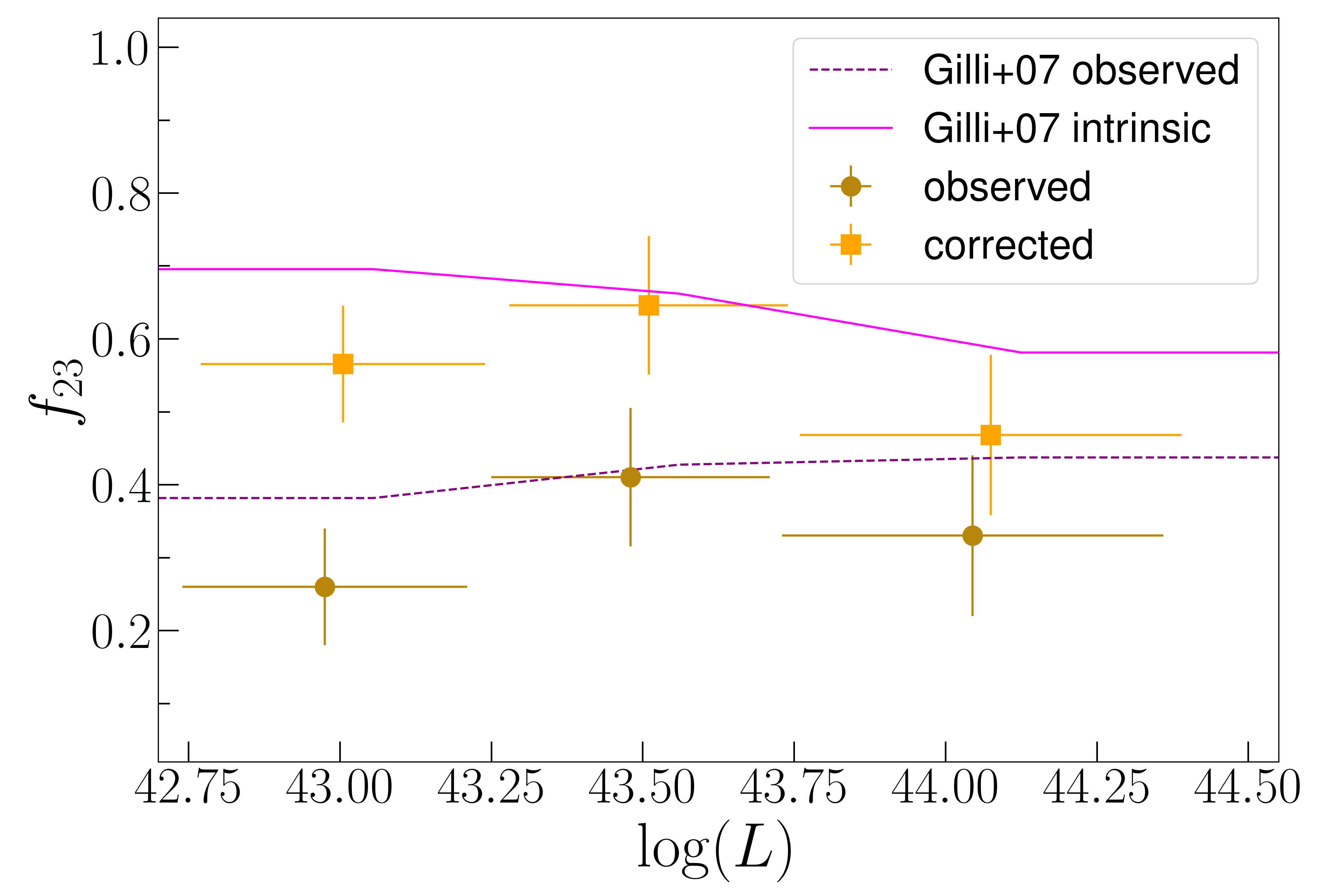}
  %\caption{Threshold: $10^{23} $cm$^{-2}$}
\end{subfigure}
\caption{Fraction of $z\sim 1.2$ obscured AGN with as a function of intrinsic 2-10 keV luminosity. Golden circles are the observed fraction, while orange squares are the ones obtained once corrected for the survey sky coverage. The second set of points is shifted by 0.05 on the log(L) axis for visual clarity. The solid magenta line represents the predictions from the \cite{Gilli07} model for the intrinsic obscured fraction; the dashed purple line shows the prediction for the observed fraction accounting for the survey sky coverage. \textit{Left}: obscured fraction derived using $N_{\rm H} > 10^{22}$\cm\ as the threshold ($f_{22}$). \textit{Right}: obscured fraction derived using $N_{\rm H} > 10^{23}$\cm\ as the threshold ($f_{23}$).}
\label{fig:f_L}
\end{figure*}

\begin{figure*}
\centering
\begin{subfigure}{.5\textwidth}
  \centering
  \includegraphics[width=.99\linewidth]{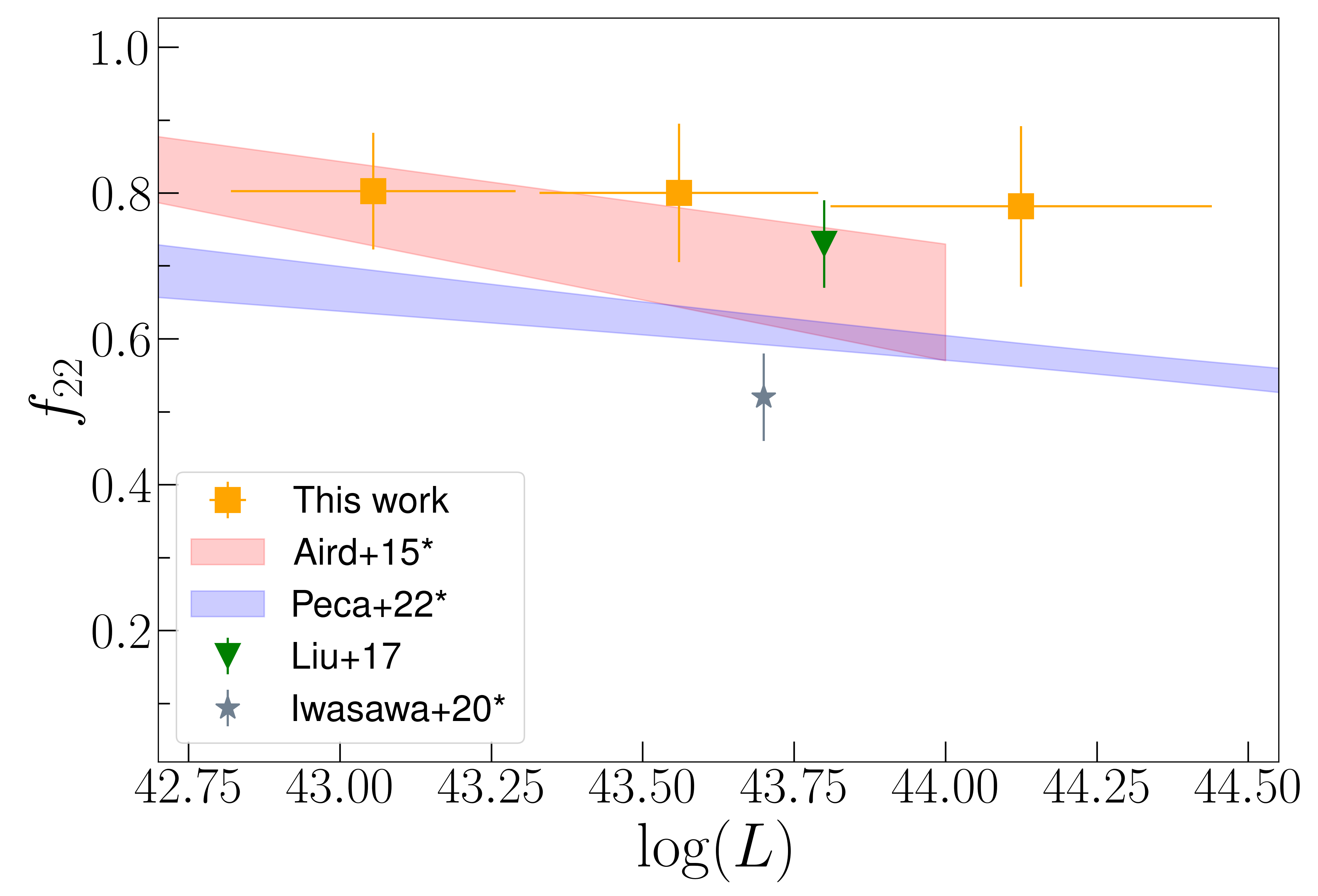}
  %\caption{Threshold: $10^{22} $cm$^{-2}$}
\end{subfigure}%
\begin{subfigure}{.5\textwidth}
  \centering
  \includegraphics[width=.99\linewidth]{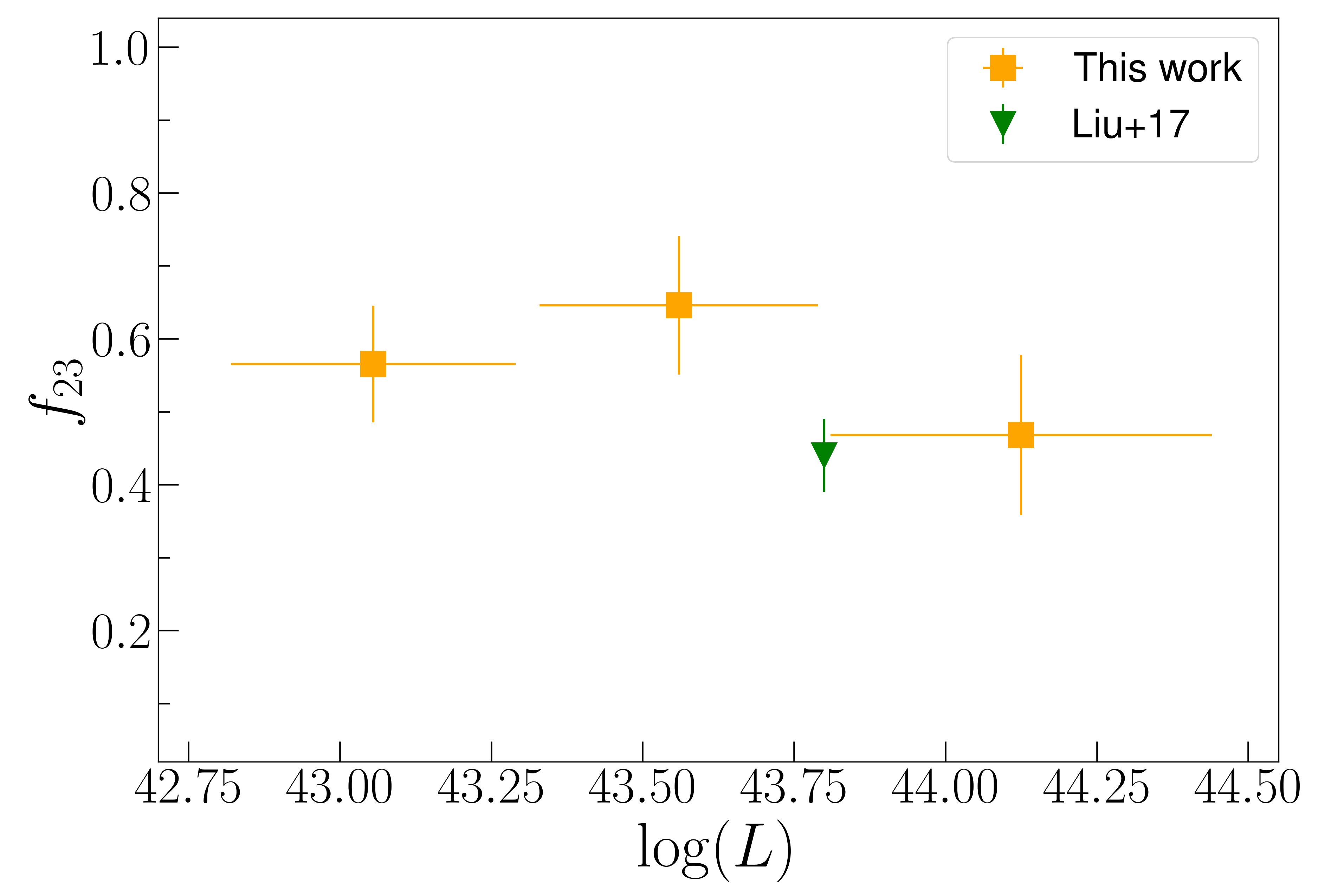}
  %\caption{Threshold: $10^{23} $cm$^{-2}$}
\end{subfigure}
\caption{Fraction of $z\sim 1.2$ obscured AGN, corrected for completeness, as a function of intrinsic 2-10 keV luminosity. Orange squares show this work results.
\textit{Left}: obscured fraction derived using $N_{\rm H} > 10^{22}$\cm\ as the threshold ($f_{22}$). This work results are compared with those of \cite{Aird15} (red shaded), \cite{Liu_17} (green triangle), \cite{Iwasawa_20} (gray star) and \cite{Peca23} (blue shade). Note: \cite{Aird15}, \cite{Iwasawa_20} and \cite{Peca23} obscured fraction consider column densities up to $10^{24}$cm$^{-2}$. \cite{Aird15} data is centered at $z\sim$1, \cite{Iwasawa_20} data is centered at $z\sim$1.35. Given the different definitions of $f_{22}$ and the redshift differences, some scatter among the results is expected.
\textit{Right}: obscured fraction derived using $N_{\rm H} > 10^{23}$\cm\ as the threshold ($f_{23}$). This work results are compared with those of \cite{Liu_17} (in green). The $f_{23}$ obscured fraction at $\log(L)\sim 44.1$ is in good agreement with that of \cite{Liu_17} at $\log(L)\sim 43.8$. \\
}
\label{fig:f_L_withothers}
\end{figure*}

We did this for each luminosity bin, starting from the fractions derived with the $p(\log(N_{\rm H}))$, and the resulting corrected fractions are shown in Table \ref{table_unc_L_f22}. In Figure \ref{fig:f_L}
we show the observed fractions, in dark gold, and the corrected fractions, in orange. We also show the magenta solid line, which is the predicted \textit{intrinsic} obscured fraction of the \cite{Gilli07} model, and the dashed purple line which is the \textit{observed} obscured fraction, given the J1030 X-ray sky coverage. 
Overall, the high uncertainties make it hard to see a clear trend of the obscured fraction with the luminosity. We can compare our results with others in the literature, with the caveat of only considering those samples of objects with a redshift similar to our ($z\sim1.2$). For the $f_{22}$ fraction, we can compare it with the work of \cite{Aird15}, for the subsample of objects that are found at a redshift $z\sim 1$, \cite{Liu_17}, for the objects found at redshift $z\sim1.2$, \cite{Iwasawa_20}, and \cite{Peca23}. For the $f_{23}$ fraction, we only have the \cite{Liu_17} data to compare with. These comparisons can be seen in Figure \ref{fig:f_L_withothers}.

The $f_{22}$ that we obtain at $\log(L)\sim 44$ are consistent with those of \cite{Aird15} and \cite{Liu_17}, while they are higher than those of \cite{Iwasawa_20} and \cite{Peca23}. Overall, the J1030 $f_{22}$ does not show a significant decline with increasing luminosity as commonly found in the literature, but, given the large error bars, it cannot be ruled out either. 
For the $f_{23}$, the estimate at $\log(L)\sim 44$ obtained in this work is consistent with the results of \cite{Liu_17}, while we lack data at different luminosities for an additional comparison. We also note that our obscured fractions at $z\sim$1.2 are on average higher than those measured by \cite{Aird15}, \cite{Iwasawa_20} and \cite{Peca23}, as expected: in said works the obscured fraction is derived as the number of objects with $ 10^{22}$cm$^{-2}<N_H<10^{24}$cm$^{-2}$ over the number of objects with $ 10^{20}$cm$^{-2}<N_H<10^{24}$cm$^{-2}$, while we considered the probability distributions of $N_{\rm H}$  from $10^{22}$cm$^{-2}$ to $10^{26}$cm$^{-2}$, that is, we included a correction for an additional population of C-thick AGN.

\subsection{Obscured fraction dependence on with redshift}\label{sec:obsc_frac_vs_z}
To investigate the redshift evolution of the obscured fraction, we performed the same analysis as the one in the Subsection \ref{sec:obsc_frac_vs_lx}, but for the three redshift bins with the same average luminosity of $\log(L)\sim 44$ (see Fig \ref{L_z}). We used the bootstrap procedure to derive, for each bin, the $f_{22}$ and $f_{23}$ and the corresponding uncertainties.

\begin{table}[ht]
\captionof{table}{Number of objects, average 2-10 keV luminosity, and fraction of AGN with $\log(N_{\rm H})>22$ ($f_{22}$) and $\log(N_{\rm H})>23$ ($f_{23}$), corrected for the completeness of the survey, in three redshift bin and relative uncertainties} % title of Table
	\centering % used for centering table
	\setlength{\tabcolsep}{3pt}
	\begin{tabular}{p{2.1cm}| p{0.4cm} p{0.95cm} p{1.7cm} p{1.7cm} } 
		\hline\hline %inserts double horizontal lines
		Bin & N & $\overline{\log(L)}$ & $f_{22}$ & $f_{23}$ \\ [0.5ex] % inserts table
		%heading
		\hline % inserts single horizontal line
        $0.8 < z < 1.6$ & 18 & 44.03 & 0.78 $\pm$ 0.11 & 0.47 $\pm$ 0.11 \\
        $1.6 < z < 2.2$ & 23 & 43.96 & 0.76 $\pm$ 0.11 & 0.61 $\pm$ 0.09 \\
        $2.2 < z < 3.2$ & 18 & 44.15 & 0.74 $\pm$ 0.11 & 0.51 $\pm$ 0.11\\
    %     \\[1ex] % [1ex] adds vertical space
		\hline 
	\end{tabular}	\label{table_unc_z} 
\\[1ex]	
%\caption*{ }
\end{table}

\begin{figure*}
\centering
\begin{subfigure}{.5\textwidth}
  \centering
  \includegraphics[width=.99\linewidth]{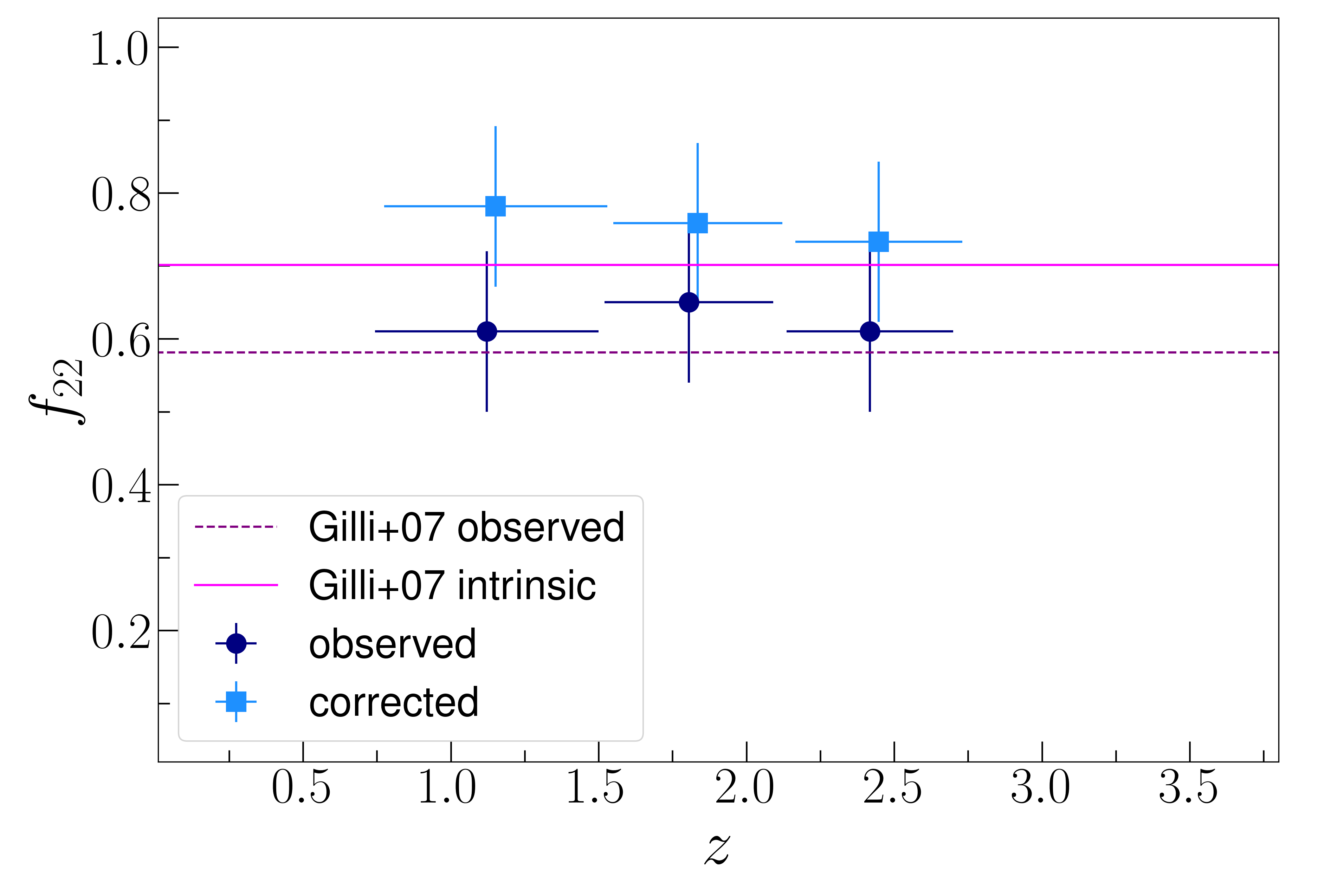}
  %\caption{Threshold: $10^{22} $cm$^{-2}$}
  \label{fig:f22}
\end{subfigure}%
\begin{subfigure}{.5\textwidth}
  \centering
  \includegraphics[width=.99\linewidth]{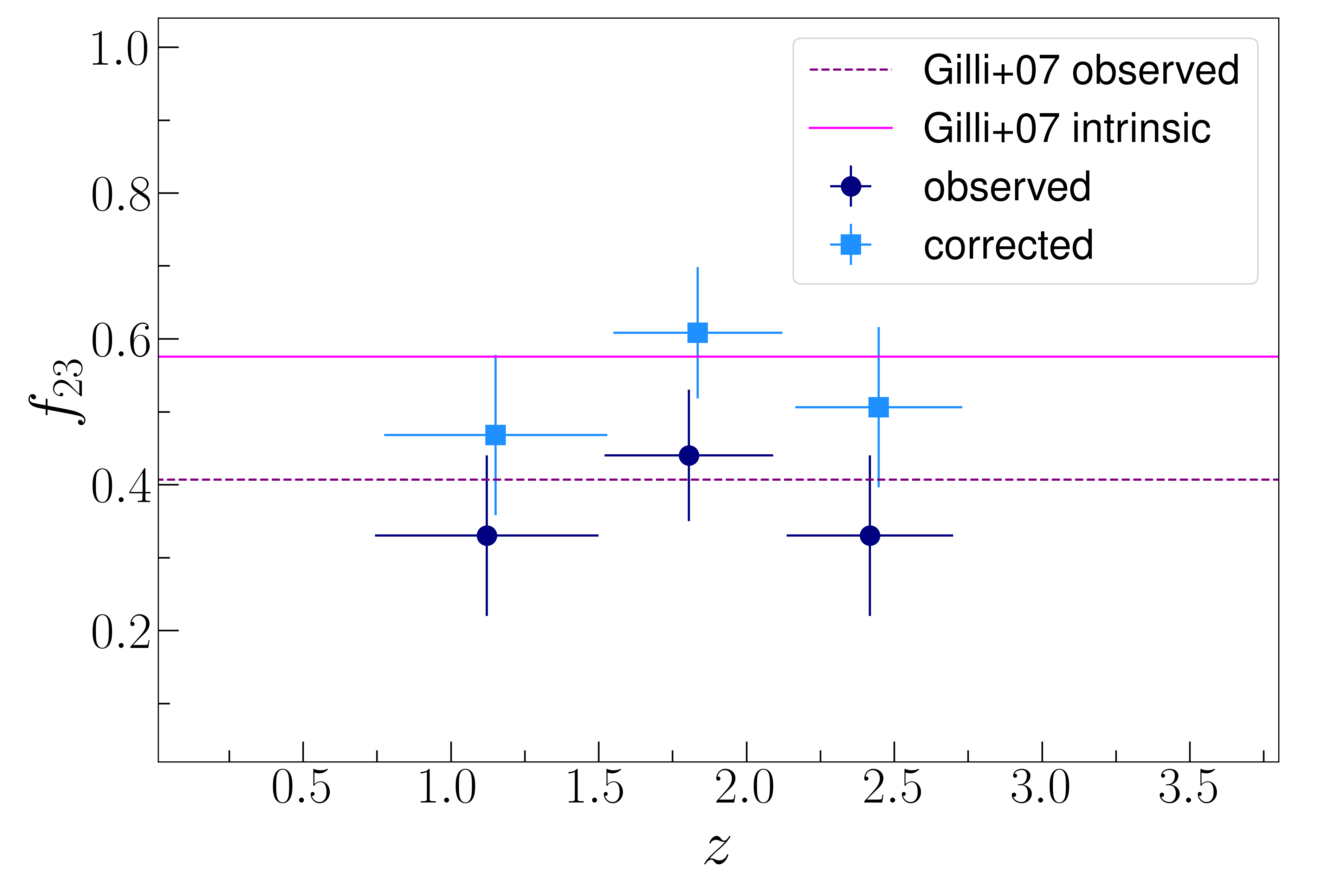}
  %\caption{Threshold: $10^{23} $cm$^{-2}$}
  \label{fig:f23}
\end{subfigure}
\caption{Fractions of $\log(L_{2-10})\sim 44$ obscured AGN as a function of redshift. Navy circles are the observed fractions, while light blue squares are those obtained once corrected for the presence of the survey sky coverage. The second set of points is shifted by 0.05 on the z-axis for visual clarity. The solid magenta line represents the predictions from \cite{Gilli07} model for the intrinsic obscured fraction; the dashed purple line is the prediction for the observed fraction once the sky coverage is taken into account. \textit{Left:} obscured fraction derived using $N_{\rm H} > 10^{22}$\cm\ as the threshold ($f_{22}$). 
\textit{Right}:  obscured fraction derived using $N_{\rm H} > 10^{23}$\cm\ as the threshold ($f_{23}$).}
\label{fig:binz_obscorr}
\end{figure*}

We then corrected the observed obscured fractions and recovered the intrinsic ones in each redshift bin using the same correction method described in Section 4.1. In Table \ref{table_unc_z}  we show the results of the corrected obscured fractions and their uncertainties in the three redshift bins. The results are shown in Figure \ref{fig:binz_obscorr}. As in the Subsection \ref{sec:obsc_frac_vs_lx}, we note that, as expected, the corrected fractions are higher than the observed ones, because the presence of a flux limit preferentially removes obscured sources from the sample.

\begin{figure*}
\centering
\begin{subfigure}{.5\textwidth}
  \centering
  \includegraphics[width=.99\linewidth]{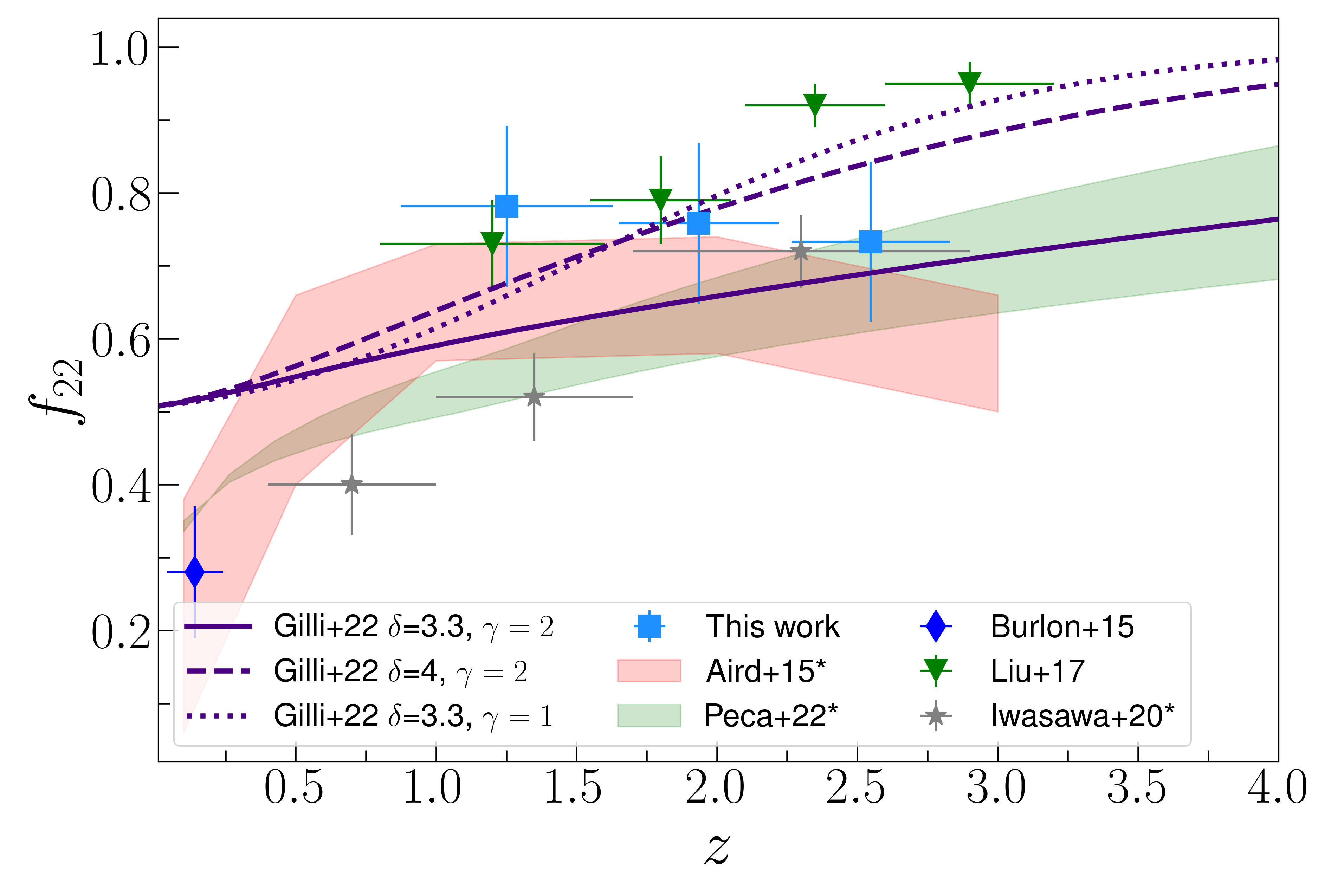}
\end{subfigure}%
\begin{subfigure}{.5\textwidth}
  \centering
  \includegraphics[width=.99\linewidth]{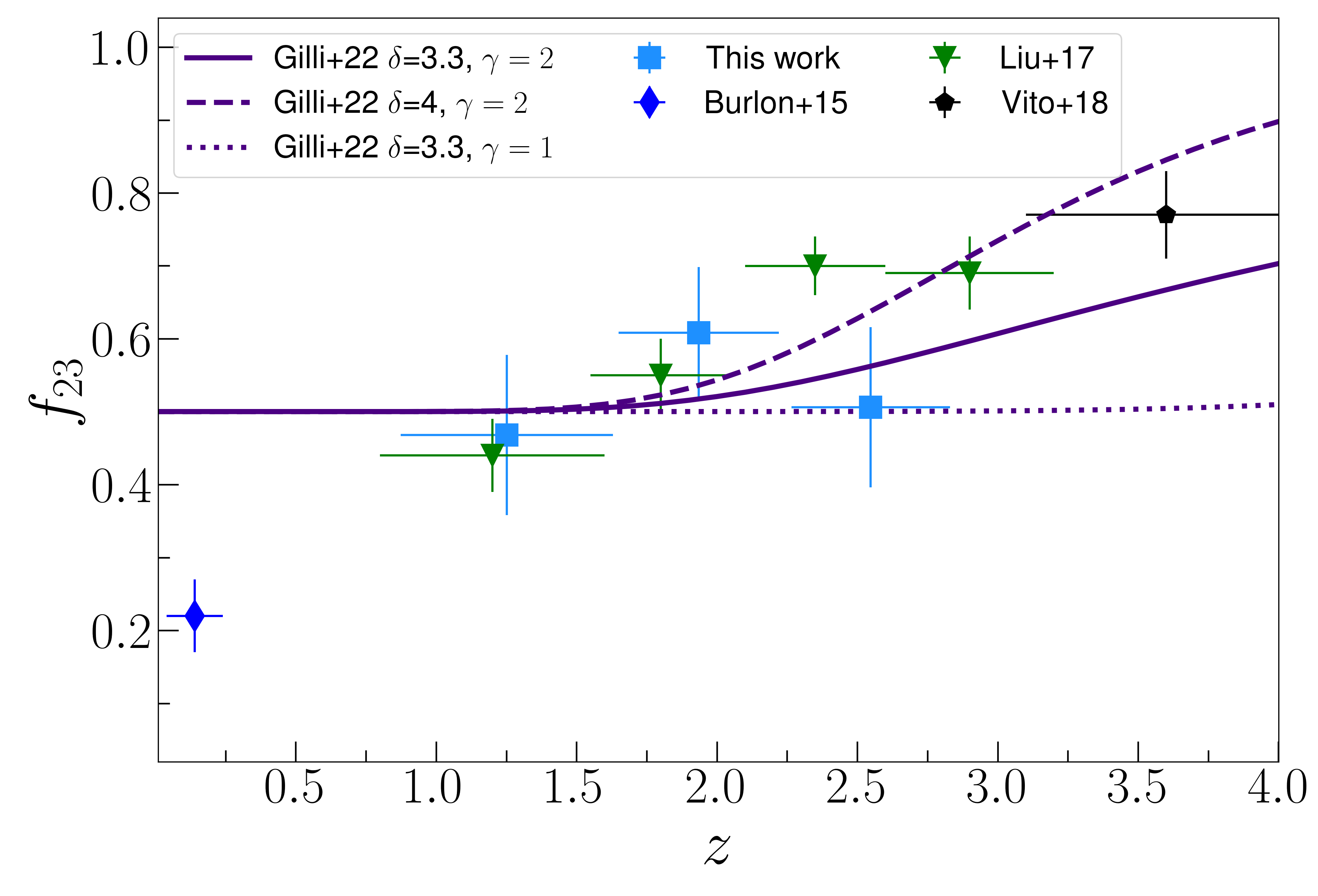}
\end{subfigure}
\caption{Fraction of $\log(L)\sim44$ obscured AGN with $N_{\rm H}>10^{22}$\cm\ ($f_{22}$) as a function of redshift. Light blue squares show this work results. The prediction of the \cite{Gilli22} model for the evolution of the obscured fraction with the redshift is shown as the indigo lines, with different styles representing different parameters of the model (see Section \ref{subsec:evolution}). 
\textit{Left}: obscured fraction derived using $N_{\rm H} > 10^{22}$\cm\ as the threshold ($f_{22}$). This work results are compared with those of \cite{Burlon11} (blue diamond), \cite{Liu_17} (green triangle), \cite{Aird15} (red shaded), \cite{Iwasawa_20} (gray star), and \cite{Peca23} (green shaded). Note: \cite{Aird15}, \cite{Liu_17}, \cite{Iwasawa_20}, and \cite{Peca23} obscured fraction consider column densities up to $10^{24}$cm$^{-2}$.  
\textit{Right:} obscured fraction derived using $N_{\rm H} > 10^{23}$\cm\ as the threshold ($f_{23}$). This work results are compared with those of \cite{Burlon11} (blue diamond), \cite{Liu_17} (green triangle), and \cite{Vito18} (black pentagon). We note that the \cite{Vito18} obscured fraction consider column densities up to $10^{25}$cm$^{-2}$.}
\label{fig:binz_withothers}
\end{figure*}

We can now compare our results with those of other works. It is important to notice that we should compare our obscured fractions with others obtained from samples with similar average luminosity. In Figure \ref{fig:binz_withothers}  we show our results (in light blue) together with those of \cite{Burlon11}, \cite{Aird15}, \cite{Liu_17}, \cite{Vito18}, \cite{Iwasawa_20}, and \cite{Peca23}, which are a representative sample of the trends in the literature. The obscured fractions in \cite{Liu_17} have been obtained in redshift ranges similar to those used in this study; the results are well consistent for the first two redshift bins, while they are more distant for the higher redshift points. The number of objects per bin in \cite{Liu_17} is roughly the same as in the J1030 sample; our uncertainties of the obscured fraction estimates are significantly higher, given the lower quality of the data and given that we took both the binomial error and the $N_{\rm H}$ uncertainties into account.

We note that \cite{Aird15}, \cite{Iwasawa_20}, and \cite{Peca23} obscured fraction do consider column densities up to $10^{24}$cm$^{-2}$. Given the different definitions of $f_{22}$ and the redshift differences, some discrepancy between the results is expected. When considering also the low-redshift results of \cite{Burlon11}, and the high redshift \cite{Vito18} estimate for $f_{23}$, there is evidence of a clear redshift trend, with bins at higher redshift showing a higher obscured fraction, both for $f_{22}$ and $f_{23}$. Overall, our results are consistent with those in the literature.

\section{Discussion}\label{sec:discussion}
In this Section, we discuss the results of our analysis and their interpretation, as well as possible limitations and biases.

\subsection{Limitations and biases}
The first limitation affecting our work is related to the sample statistics. Although the J1030 \textit{Chandra} survey is one of the deepest X-ray surveys, about 30\% of the objects have less than 30 net counts, which affects our ability to derive accurate parameters from the spectral fit. Furthermore, the progressive degradation of the \textit{Chandra} detectors affects the soft X-ray response in a non-negligible way. As the spectral shape at low energies is more informative to distinguish between different levels of obscuration, our ability to derive reliable $N_{\rm H}$ estimates is reduced. The low sensitivity at low energies also significantly skews the $N_{\rm H}$ probability distributions towards low $N_{\rm H}$ values, even when a significant probability peak around a certain value is found (see e.g. XID 77 and XID 116 in Fig.4). These uncertainties clearly affect the accuracy with which we can estimate the obscured AGN fractions, compared with surveys with longer exposures and with surveys where observations have been carried out in earlier years of the \textit{Chandra} satellite life.

Another source of uncertainty comes from the fact that $44\%$ of the objects in our sample only have a photometric redshift estimate. In the fit procedure, we considered the redshift as a fixed parameter. 
However, the errors on the photometric redshifts can be significant. This affects again the accuracy of the $N_{\rm H}$ estimates. Furthermore, in the obscured AGN fraction analysis, some objects that fall in a given luminosity-redshift bin might actually belong to other, adjacent bins. Observational campaigns aimed at improving the spectroscopic redshift completeness of the J1030 \textit{Chandra} sample are being planned. 

We measured the obscured AGN fractions in different bins of X-ray luminosity and redshift. The main source of errors on these fractions is related to (i) the limited sample statistics in each luminosity-redshift bin, and (ii) the uncertainties in the column density estimate of each source. Given that the uncertainties we derive from the Wilson score values are in the $\sim$0.05-0.10 range, compared to total uncertainties derived from the bootstrapping procedure of $\sim 0.11$, we can say that the first contribution is generally more significant than the second one. When compared with the results obtained from other surveys, our uncertainties are significantly higher. This depends on the fact that, in general, previous studies do not take \textit{both} sources of uncertainties into account, on the lower data quality of our X-ray spectra when compared with other samples (e.g. \citealt{Liu_17}), and on the higher statistics of other surveys.

The uncertainties in our obscured fraction estimates are such that we do not have clear evidence of a redshift or a luminosity trend with the J1030 data alone (see Figures~\ref{fig:f_L_withothers} and \ref{fig:binz_obscorr}). At the same time, as shown in Fig.~\ref{fig:binz_withothers}, our results are generally consistent with those in the literature for AGN with similar luminosities and at similar redshifts. Furthermore, when combined with samples of X-ray selected AGN covering a broader range of redshifts, our results follow the general literature trends, where the obscured AGN fraction gets higher towards higher redshifts.

Another bias that might be affecting our results is the \textit{classification bias}: when an object has a small number of counts, heavily-obscured objects can be misclassified as mildly-obscured ones \citep[for more details, see][]{Brightman12, Lanzuisi18}. The low-luminosity objects are the ones with a smaller number of counts, therefore the most affected by this bias. In terms of the obscured fraction trend with the luminosity, this means that we are probably underestimating the obscured fraction in the first luminosity bin, which might be preventing us from seeing a clear trend. 

\subsection{Evolution of the obscured AGN fraction}\label{subsec:evolution}

While the decrease of the obscured AGN fraction with luminosity is generally interpreted in the framework of the so-called receding-torus models \citep{Lusso13, Ricci17}, the physics behind the increasing trend of the obscured fraction with redshift is not completely understood. 

We compared our results with the model recently proposed in \cite{Gilli22}. In that work, the authors argue that the evolution of the obscured AGN fraction is produced by the increasing density of the ISM in the host galaxies, and give an analytic model for that. In Figure \ref{fig:binz_withothers} we show the predictions of the baseline model of \cite{Gilli22} as the solid indigo lines. The other lines reflect different assumptions in the model parameters that we will now discuss. Considering the baseline model, we see that there is a good agreement for $f_{23}$, while for $f_{22}$ our values are higher than the prediction, although consistent at a 1.5$\sigma$ level. Our measurements are generally in better agreement with the model curves than the measurements of \cite{Liu_17}, who found larger obscured AGN fractions at $z>2$. We recall, however, that the model curves from \cite{Gilli22} are an example of how the increased ISM density may provide a good representation of the observed trend, but they were not derived through best-fit procedures to any specific dataset. Here we explore the parameter space of that model further, trying to determine, for instance, how the ISM properties should change with redshift to reproduce the steeper trend observed by \cite{Liu_17}.

By considering a number of tracers for the total mass and volume of the ISM in galaxy samples at different redshifts, mainly from ALMA, and simple assumptions on the gas density profiles, \cite{Gilli22} measured the cosmic evolution of the ISM column density towards the nuclei of massive galaxies. This was parameterized as   
\nhism$\propto(1+z)^\delta$. They also assumed that the ISM is composed of individual gas clouds with surface densities and radii distributed as a Schechter function and that the characteristic cloud surface density $\Sigma_{c,*}$ may evolve with redshift as $(1+z)^\gamma$. The redshift evolution of the ISM-obscured AGN fraction above a given \nhism\ threshold depends on both $\delta$ and $\gamma$ (see Eqs.40 and 41 in \citealt{Gilli22}). Broadly speaking, a rapid increase of the total column density with redshift would imply a correspondingly rapid increase of the obscured AGN fraction. This increase is nonetheless softened if ISM clouds are significantly denser at earlier cosmic epochs, as fewer clouds would then be needed to reproduce the same total gas density, reducing, in turn, the chances that galaxy nuclei are hidden by one of those. The baseline model in \cite{Gilli22} assumed $\delta=3.3$, as driven by the results from ALMA observations, and $\gamma=2$, which, when combined with the obscuration from a small-scale component (i.e. the torus) was found to produce $f_{22}$ and $f_{23}$ trends in good agreement with the observations. Clearly, the uncertainties on $\delta$ and $\gamma$ are large, as we still have limited knowledge of the overall ISM properties of distant galaxies. In Fig. \ref{fig:binz_withothers} we show the expected trends for $f_{22}$ and $f_{23}$ when first increasing $\delta$ and then decreasing $\gamma$, leaving all the other model parameters unchanged.
A faster increase of the total ISM density with redshift ($\delta=4$) is needed to explain
the steep trend observed by \cite{Liu_17} for $f_{22}$ and $f_{23}$ and the large $f_{23}$ value  measured by \cite{Vito18} at $z\sim3.6$. On the other hand, interestingly, a milder evolution of the characteristic gas surface density of ISM clouds ($\gamma=1$) would only explain the steeper trend in $f_{22}$ but not in $f_{23}$. This is because, below $z\sim4-5$ the distribution of cloud surface densities would be rich of clouds with $\Sigma_{c,*}>10^{22}$\cm, but still short of high-density clouds with $\Sigma_{c,*}>10^{23}$\cm. It is only at $z\sim6$ and above that $\Sigma_{c,*}$ would increase enough to return significant fractions of very dense clouds.

To summarize, current measurements of the obscured AGN fractions as a function of cosmic time, including ours, are in agreement with an evolving ISM model in which the total gas column density of massive galaxies evolves as fast as \nhism$\propto(1+z)^{3.3-4}$, and in which the individual gas clouds become progressively denser towards early epochs [$\Sigma_{c,*}\propto(1+z)^2$]. Such a scenario will likely be tested soon with improved accuracy by new ALMA observations.

\subsection{Compton-thick AGN}
Our work only considers the X-ray spectral fitting as an obscuration diagnostic. Because of this, it is likely that we are not able to correctly characterize heavily obscured objects, especially Compton Thick (CT) AGN, which will also tend to have a small number of counts. In addition to this, absorption models like the one we used (\textit{phabs}) do not work well in a very high column density regime.
We find 8 objects with a nominal $N_{\rm H}$ higher than $10^{24}$cm$^{-2}$ out of 243, which means that we have a CT fraction of $3.3\%$. If we consider the $p(\log(N_{\rm H}))$ and sum all the fractions with $N_{\rm H} > 10^{24}$cm$^{-2}$, we get an observed fraction $f_{24} = 3\%$, close to the one we get from nominal values. This value is smaller than the $\sim8\%$ CT fraction that is found by \cite{Liu_17} for the \textit{Chandra} Deep Field South. However, in that work, the authors use additional criteria other than the X-ray spectral fitting to determine if a source is Compton Thick. In \cite{Lanzuisi18}, instead, where the only diagnostic is again the X-ray spectral analysis, the CT fraction found in the COSMOS \textit{Chandra} survey was $2.2\%$, similar to our result.

Based on these previous results, it is therefore likely that if additional multi-band diagnostics were implemented, we would get a higher number of CT objects. Therefore, the CT fraction that we get is to be considered as a lower limit for the intrinsic one.

\section{Conclusions}\label{sec:conclusions}
In this work, we analyzed the X-ray spectra of the 243 extragalactic sources of the J1030 \textit{Chandra} catalog and used the results to derive the obscured fraction of AGNs at different redshift and luminosities. Here we outline the main results of our work and future perspectives. 
\begin{itemize}
    \item We fitted the Chandra X-ray spectra with absorbed power-laws, and checked for the presence of the Fe K$\alpha$ line and a soft excess. We could use spectroscopic redshift information for 44\% of the sample, while we relied on photometric redshift estimates for the rest. For 7 objects with a photometric redshift only, we were able to refine the redshift estimate via X-ray spectroscopy. The best fit spectral parameters derived for the whole sample are available at the J1030 website \footnote{\url{http://j1030-field.oas.inaf.it/chandra_1030}}.
    \item We measured the obscured fractions $f_{22}$ and $f_{23}$ (i.e. the fraction of AGN with $N_{\rm H} > 10^{22}$cm$^{-2}$ and $10^{23}$cm$^{-2}$, respectively) using the full column density probability distributions derived from the spectral fits $p(\log(N_{\rm H}))$. We measured $f_{22}$ and $f_{23}$ in three redshift bins ($0.8 < z < 1.6, 1.6 < z < 2.2 $ and $
    2.2 < z < 2.8$) for AGN with $\log(L) \sim 44$, and in three luminosity bins ($42.8 < \log(L_{2-10keV}) < 43.3, 43.3 <\log(L_{2-10keV}) < 43.8,
    $ and $ 43.8 < \log(L_{2-10keV}) < 44.5$), for AGN at $z\sim$ 1.2. We corrected these \textit{observed} fractions for the sky coverage of the survey and derived accurate measurement errors through a bootstrapping procedure that accounts for both the finite size of the sample and the uncertainties on the $N_{\rm H}$ estimates.
    \item We measured average values of $f_{22}\sim0.7-0.8$ and $f_{23}\sim0.5-0.6$. While these average values are in broad agreement with those in other works \citep{Aird15,Liu_17}, we did not see clear trends with luminosity or redshift, as opposed to what is often found in the literature. This might, at least partially, depend on residual, uncorrected biases, and/or on the limited dynamical range in luminosity and redshift spanned by our data. Nonetheless, when combined with measurements performed in the local Universe, our data point to an increase of the obscured AGN fractions with redshift, in agreement with other findings.
    \item We finally considered predictions from recent analytic models that ascribe the redshift evolution of the obscured AGN fraction to the increased density of the ISM in high-z hosts, which adds significant obscuration to that of the pc-scale 'torus' \citep{Gilli22}. When combined with literature measurements, our results favor a scenario in which the total ISM column density grows with redshift as \nhism$\propto(1+z)^{3.3-4}$, and in which the characteristic surface density of individual gas clouds in the ISM evolves as $\Sigma_{c,*}\propto(1+z)^2$.

\end{itemize}

To gain a deeper understanding of nuclear obscuration at different cosmic epochs, and as a function of the various AGN physical properties, large object samples are needed, that would go significantly beyond those available from current X-ray probes. What is believed to be the bulk of the AGN population (low-luminosity, possibly obscured objects) is now partly missed at medium-high redshift values, and completely lost beyond redshift $z \sim6$. Next-generation X-ray imaging surveys, such as those proposed with the Survey and Time-domain Astrophysical Research eXplorer (\textit{STAR-X}\footnote{\url{http://star-x.xraydeep.org/}}), a Medium Explorer mission selected by NASA for Phase A study, the Advanced X-ray Imaging Satellite (\textit{AXIS}, \citealp[]{Mushotzky19, Marchesi20}), a probe-class mission proposed to NASA, and the L-class mission \textit{Athena} under scrutiny at ESA \citep{Nandra13}, would offer new opportunities to detect and characterize highly obscured sources. These observatories are expected to discover a few thousands of heavily obscured ($N_{\rm H}>10^{23}$~cm$^{-2}$) AGN at $z>3$, shedding light on the overall growth of SMBHs before cosmic noon.

\begin{acknowledgements}
We thank the referee for their detailed and constructive suggestions. We acknowledge financial contribution from the agreement ASI-INAF n. 2017-14-H.O.
\end{acknowledgements}

\bibliographystyle{aa}
\bibliography{bibl}

\begin{appendix}

\section{Fitting procedure: background}\label{sec:app_bkg_fit}
In this Appendix we describe our fitting procedure for the background spectra of our objects. In order to have the most reliable estimates of the parameters we want to fit, we decide to simultaneously fit the source and the background.

We have extracted local background spectra, but we do not want to model the background locally as (i) we want to minimize dependencies on sharp background variations, (ii) the number of counts in background spectra extracted in small regions around each source can be very small and therefore the parameters uncertainties very high. We want therefore to characterize the whole background and then use the fitted background model to simultaneously fit each source with its local background, adding a "background normalisation" parameter to the source fit to rescale the normalisation to that of the local background. \\We selected three regions, all centered in the center of the field: one circular region of radius 3 arcmin, one annulus with 3 and 6 arcmin as radii and an annulus with 6 and 9 arcmin as radii. In each region we excluded circular regions of 5 arcsec radius around all the X-ray detected sources. We extracted the spectrum of each region in the energy range 0.8-7 keV, which is the one we use to fit the sources. We modeled each background spectrum with a power-law and four Gaussian lines, following the model used by \cite{Fiore12} for the \textit{Chandra} Deep Field South survey. Following the same model, we also tried to (i) add a second power-law component, and (ii) add a thermal component, but both turned out to be non significant. We therefore excluded these components from the final background shape, which ends up being composed of a power-law and four Gaussians.

\begin{figure*}
    \centering
    \begin{subfigure}[t]{0.49\textwidth}
        \centering
        \includegraphics[width=1.\linewidth]{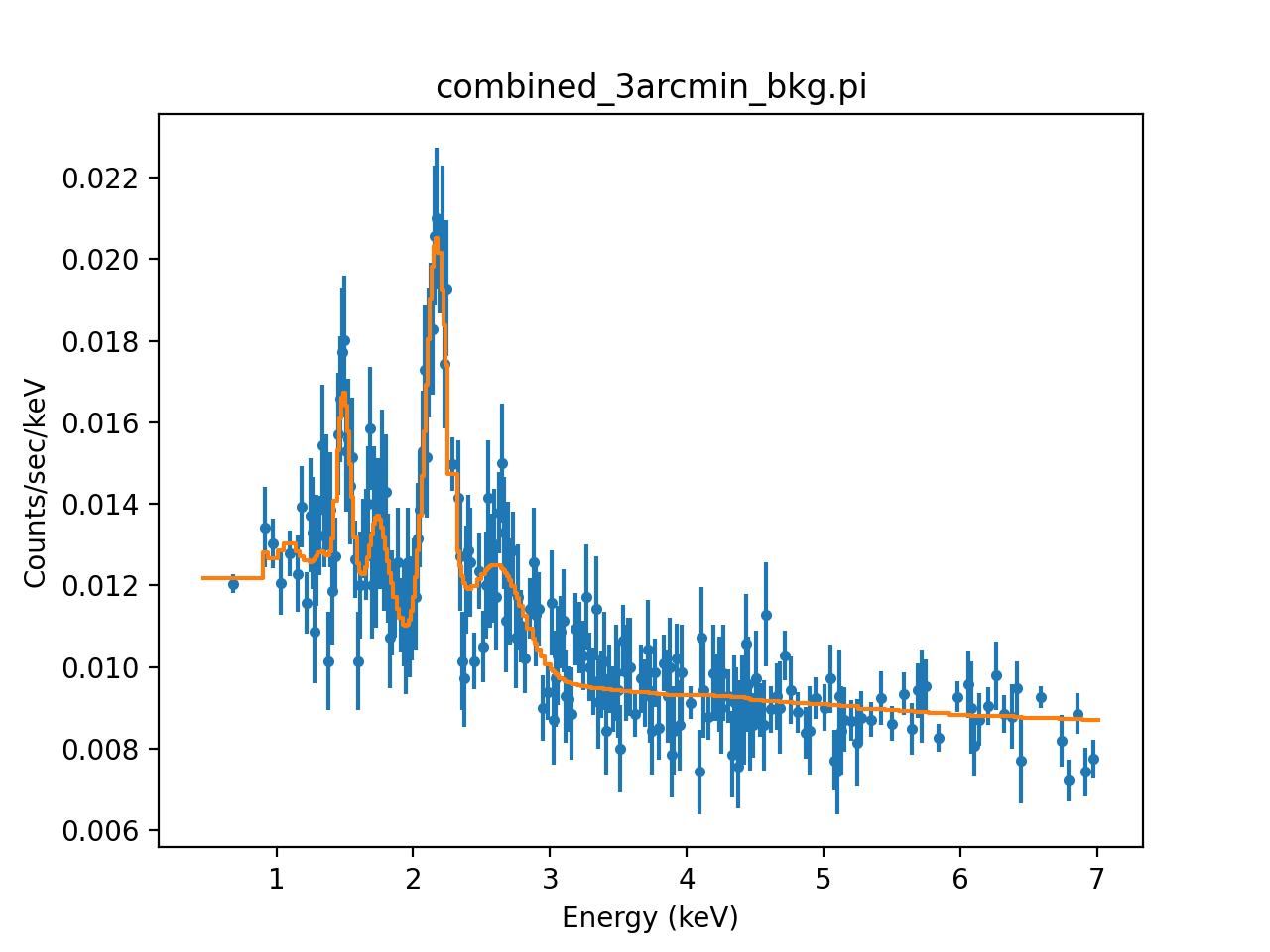} 
    \end{subfigure}
    \hfill
    \begin{subfigure}[t]{0.49\textwidth}
        \centering
        \includegraphics[width=1.\linewidth]{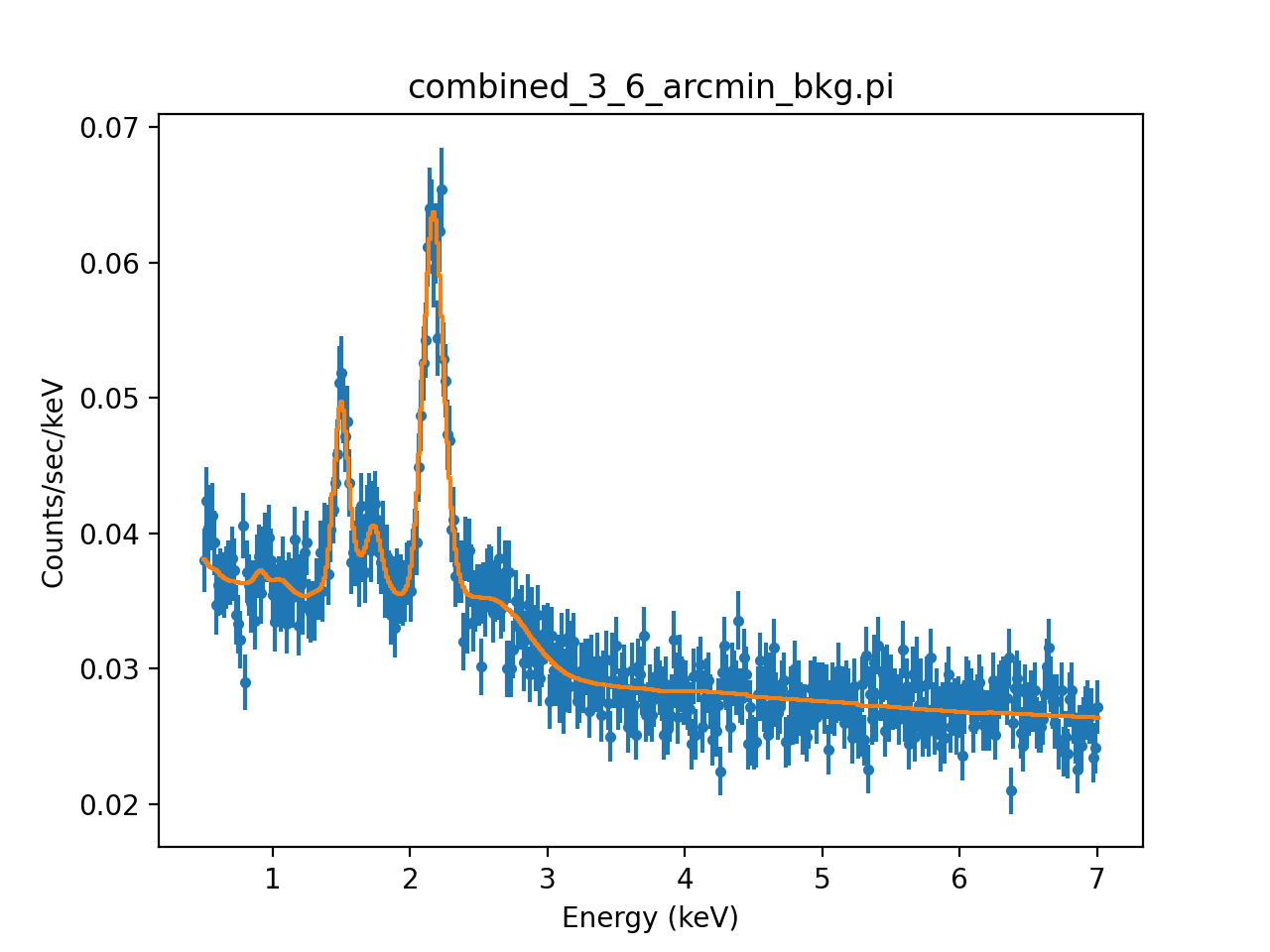} 
    \end{subfigure}

    \begin{subfigure}[t]{0.49\textwidth}
    \centering
        \includegraphics[width=1.\linewidth]{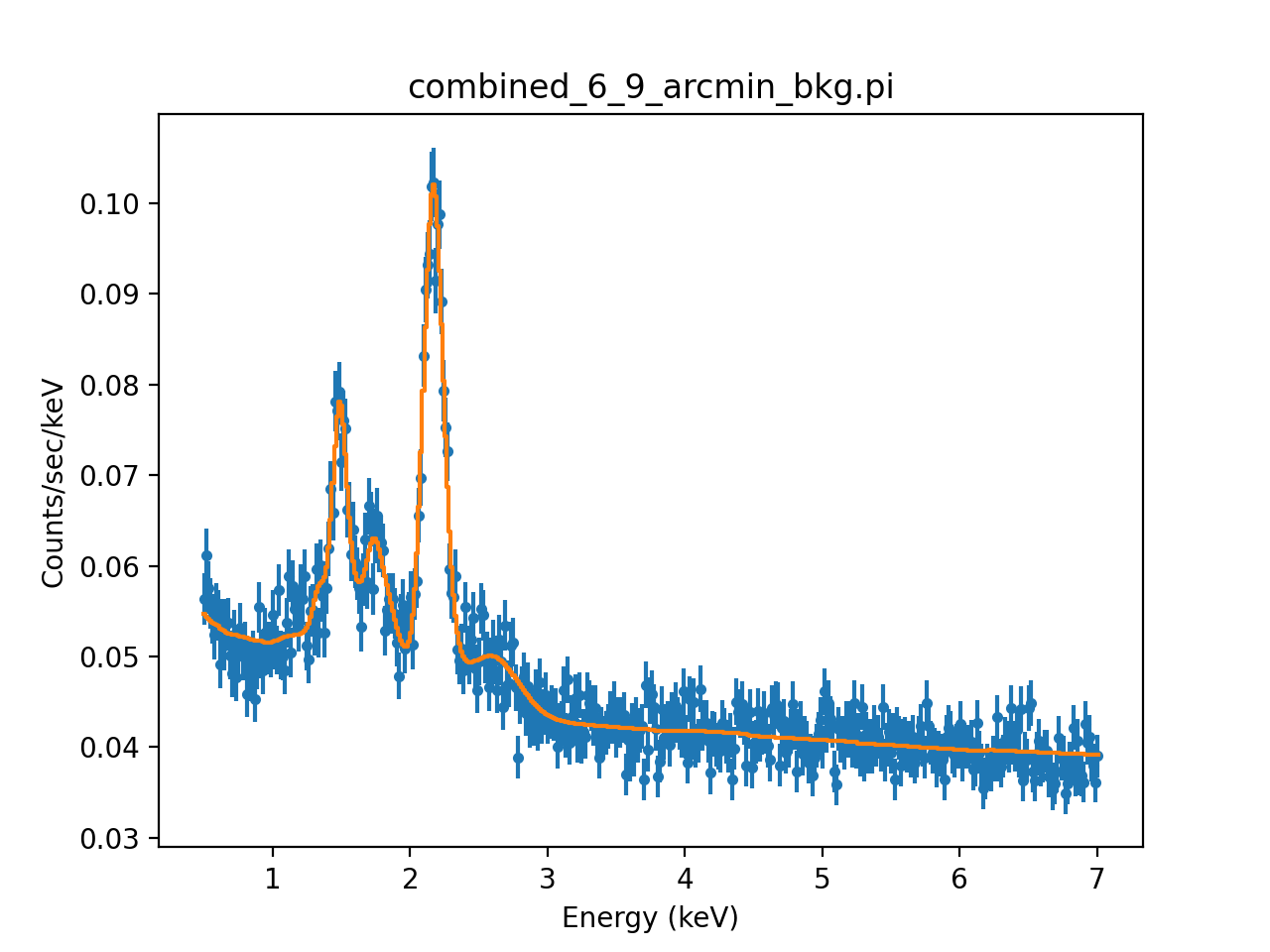} 
    \end{subfigure}
    \caption{Spectrum (blue) and best fit (in yellow) of the background spectra for different regions of the field: (a) 3' circle; (b) annulus of radii 3' and 6'; (c) annulus of radii 6' and 9'}\label{fig:bkg}
\end{figure*}

In Figure~\ref{fig:bkg} we can see the spectra and the resulting best fit.
Given the best fit parameters of the modeled background, we used them as "freezed" parameters in the source+background fit analysis, only adding a multiplicative constant as a free parameter to re-scale the background spectrum to the one of each object.

\end{appendix}
\end{document}